\documentclass[fleqn,usenatbib]{mnras} 

\usepackage{newtxtext,newtxmath,bm}
\usepackage[T1]{fontenc}
\usepackage[dvipsnames]{xcolor}
\usepackage{soul}

\usepackage{CJKutf8} 

\DeclareRobustCommand{\VAN}[3]{#2}
\let\VANthebibliography\thebibliography
\def\thebibliography{\DeclareRobustCommand{\VAN}[3]{##3}\VANthebibliography}

\usepackage{graphicx}	
\usepackage{amsmath}	

\newcommand{\im}[1]{\textcolor{black}{#1}}

\newcommand{\haiyang}[1]{\textcolor{black}{#1}}
\newcommand{\dong}[1]{\textcolor{black}{#1}}

\title{Hydrodynamical Simulations of Circumbinary Accretion: Balance between Heating and Cooling}

\author[H. Wang et al.]{
Hai-Yang Wang \begin{CJK*}{UTF8}{gbsn}(王海洋)\end{CJK*},$^{1,2}$\thanks{cgmgalaxy0721@gmail.com}
Xue-Ning Bai \begin{CJK*}{UTF8}{gbsn}(白雪宁)\end{CJK*},$^{3}$\thanks{xbai@tsinghua.edu.cn}
Dong Lai \begin{CJK*}{UTF8}{gbsn}(赖东)\end{CJK*}$^{4}$\thanks{dong@astro.cornell.edu}
Douglas N. C. Lin \begin{CJK*}{UTF8}{gbsn}(林潮)\end{CJK*}$^{3,5}$\thanks{lin@ucolick.org}
\\
$^{1}$Fudan University, Department of Physics, Shanghai 200433, China\\
$^{2}$Department of Applied Mathematics and Theoretical Physics, University of Cambridge, Wilberforce Road, Cambridge CB3 0WA, UK\\
$^{3}$Institute for Advanced Study and Department of Astronomy, Tsinghua Univeristy, Beijing 100084, China\\
$^{4}$Department of Astronomy, Center for Astrophysics and Planetary Science, Cornell University, Ithaca, NY 14853, USA\\
$^{5}$Department of Astronomy \& Astrophysics, University of California, Santa Cruz, CA 95064, USA
}

\date{Accepted XXX. Received YYY; in original form ZZZ}
\pubyear{2023}

\graphicspath{{./}{Figures/}}

\begin{document}
\label{firstpage}
\pagerange{\pageref{firstpage}--\pageref{lastpage}}
\maketitle

\begin{abstract}

Hydrodynamical interaction \dong{in} circumbinary discs (CBDs) plays a crucial role \dong{in various astrophysical systems, ranging from young stellar binaries to supermassive black hole binaries in galactic centers}. 
\dong{Most previous simulations of binary-disc systems have} adopted locally isothermal equation of state.
In this study, we use the grid-based code \texttt{Athena++} to conduct a suite of two-dimensional viscous hydrodynamical simulations of \dong{circumbinary accretion} on a cartesian grid, \dong{resolving the central cavity of the binary}. 
The gas thermodynamics is treated by thermal relaxation towards an equilibrium temperature \dong{(based on the constant$-\beta$ cooling ansatz, where $\beta$ is the cooling time in units of the local Keplerian time)}.
Focusing on equal mass, circular binaries in CBDs with \dong{(equilibrium)} disc aspect ratio $H/R=0.1$, we find that the cooling of the disc gas \dong{significantly} influences the binary orbital evolution, accretion variability, and CBD morphology, \dong{and the effect depends sensitively on } the \dong{disc viscosity prescriptions}. When adopting \dong{a} constant kinematic viscosity, \dong{a finite cooling time ($\beta \gtrsim 0.1$)} leads to binary inspiral as opposed to outspiral and the CBD \dong{cavity} becomes more symmetric. When adopting \dong{a dynamically varying} $\alpha-$viscosity, \dong{binary} inspiral only occurs within a narrow range of cooling time \dong{(corresponding to $\beta$ around 0.5)}.
\end{abstract}


\section{Introduction}
\label{section:introduction}


Circumbinary discs (CBDs) can be present around a wide range of \dong{binary} systems, \dong{and} play a crucial role \dong{in their} evolution.
In the star formation context, CBDs can form as a natural product \dong{of disc or core fragmentations and dynamical interactions}~\citep{1986ApJS...62..519B,1994MNRAS.269L..45B,1994MNRAS.271..999B,2008ApJ...681..375K}. While the CBDs \dong{affect} binary evolution, binaries can also leave observable imprints on the CBDs; \dong{examples include} Class I/II systems GG Tau, DQ Tau and UZ Tau E~\citep{1994A&A...286..149D,1996A&AS..117..393B,1997AJ....113.1841M} \dong{and Class 0 system} L1448 IRS3B~\citep{2016Natur.538..483T}. Moreover, the discovery of \dong{circumbinary} planets~(e.g., \citealt{2018MNRAS.480.3800H} and references therein) triggered substantial interest in studying planet formation and migration in CBDs, which are closely connected to binary-disc interactions~\citep{2014A&A...564A..72K,2015A&A...581A..20K,2017MNRAS.465.4735M,2018A&A...616A..47T,2021A&A...645A..68P}. 



CBDs are also expected to form around binary supermassive black holes (SMBHs) following galaxy mergers. They may play a key role in the orbital evolution of SMBH binaries, especially for orbital separations between $\sim$ 0.01 and $\sim$ 1 pc. At larger separations, dynamical friction by gas and star and stellar ``loss-cone'' scattering are  effective in driving the orbital decay of binary SMBHs ($\gtrsim$ 1 pc, e.g.,  \citealt{2005ApJ...630..152E}), while at small separations the emission of gravitational waves is effective ($\lesssim$ 0.01 pc, e.g.,  \citealt{2003ApJ...583..616J,2003ApJ...590..691W,2008MNRAS.390..192S,2017MNRAS.464.3131K}). The orbital evolution of SMBH binaries in the intermediate range, known as the "final parsec problem", has been a central topic of theoretical studies of CBDs \citep{1980Natur.287..307B,2002ApJ...567L...9A,2008ApJ...672...83M,2009ApJ...700.1952H}.

\dong{Without accretion, a binary is expected to lose angular momentum to its CBD}  through resonantly excited trailing density waves~\citep{1979ApJ...233..857G}. Some \dong{earlier} numerical simulations \dong{have claimed a } negative torque exerted on the binary~(e.g.,  \citealt{2008ApJ...672...83M})\dong{, but this likely reflects the ``transient'' state of the simulation, as the flow has not reached a quasi-steady state. In general,} the CBD is not only the source of torque acting on the binary. The accretion streams penetrating the disc cavity and the mini-discs formed around each \dong{binary} component can also provide non-negligible amount of gravitational torque, as well as the torque stems from direct accretion onto the binary. \dong{All these torque components must be included and time-averaged in simulations of sufficiently long durations in order to determine the long-term orbital evolution of the binary.}


Recently, significant progress has been made towards more comprehensive understandings of CBD physics through numerical simulations, most of which are conducted in 2D viscous hydrodynamics (see \citealt{2022arXiv221100028L} for a review). It \dong{has been recognized} that reaching a quasi-steady state of the \dong{flow} is needed before applying diagnostics to calculate binary orbital evolution (e.g., \citealt{2017MNRAS.466.1170M,2019ApJ...871...84M,2020ApJ...889..114M}). As the most extensively studied case, equal-mass, circular binaries in CBDs with disc aspect ratio $h=H/r=0.1$ \dong{were found to} experience orbital expansion~(e.g., \citealt{2017MNRAS.466.1170M,2019ApJ...871...84M}). This result has been \dong{confirmed by other in }the 2D and 3D simulations~\citep{2019ApJ...875...66M}, and is largely insensitive to the choices of sink prescriptions~\citep{2021ApJ...921...71D} and boundary conditions
~\citep{2020ApJ...889..114M} \dong{for the binary components}.
Later, it was found that the dynamics of the CBDs and the torque exerted on the binary can sensitively depend on parameters related to the binary and disc properties: Binary will experience a transition from outspiral to inspiral when the disc aspect ratio $h\lesssim0.04$,~\citep{2020A&A...641A..64H,2020ApJ...900...43T,2022MNRAS.513.6158D}, or when the mass ratio $q\lesssim0.05$~\citep{2020ApJ...889..114M,2020ApJ...901...25D}. The orbit of the binary can also expand or shrink depending on the eccentricity of the binary~\citep{2016ApJ...827...43M,2019ApJ...871...84M,2020ApJ...889..114M,2021ApJ...914L..21D,2021ApJ...909L..13Z} and the viscosity in the CBD~\citep{2020ApJ...901...25D,2022A&A...660A.101P,2022MNRAS.513.6158D}.

\dong{All the studies mentioned above have adopted the locally isothermal equation of state, i.e.}, assuming the disc temperature is a fixed function of position \dong{such that} the disc aspect ratio $h$ \dong{is a constant}. There are only a handful of works \dong{that consider} different \dong{gas thermodynamics}. Most of \dong{these focused on other aspects of the problem, such as the role} of self-gravity \citep{2021MNRAS.507.1458F}, the evolution of disc cavity and modeling of specific binary systems \citep{2018A&A...616A..47T,2019A&A...627A..91K,2022A&A...664A.157S}. Cooling in the CBDs is also included in \dong{some GRMHD} simulations, in the regime near binary merger~(e.g., \citealt{2012ApJ...755...51N,2018ApJ...865..140D,2021ApJ...922..175N}). \im{The impact of different temperature profile on the evolution of black hole binaries embedded in the AGN disks has been studied by \citet{2022ApJ...928L..19L}.}

In our previous work \citep{paper0}, we studied the effect of dynamical cooling (\dong{use the $\beta$ prescription; see below}) on disc morphology, binary accretion variability and binary orbital evolution \dong{using cylindrical-grid simlations with excised binary cavity}. We found that with a longer cooling time, the disc becomes more symmetric and the accretion variability is gradually suppressed. \dong{We also found that the relationship} between the rate of angular momentum transport and the cooling time exhibits a ``v''-shaped structure. Namely, the rate of angular momentum transport first decreases to a minimum value then increases as the cooling time increases. \dong{However, we can only} explore a very limited parameter space using cylindrical-grid simulations with an excised central cavity \dong{-- such simulations} do not capture the mini-discs and accretion streams around individual binary components.

In this paper, we aim at a comprehensive study on \dong{how the gas thermodynamics affects binary-CBD interactions}. Following \citet{paper0}, we consider an equal mass binary on a fixed circular orbit, and treat the disc thermodynamics through dynamical cooling, relaxing disc temperature to a given temperature profile over a cooling time $t_{\rm{cool}}$ \citep{2001ApJ...553..174G}. A significant improvement is that we resolve the mini-discs around each binary component by carrying out simulations in the cartesian geometry. In doing so, we \dong{find that in addition to the cooling time}, different temperature and viscosity prescriptions in the central cavity and the mini-discs also have a significant impact on the CBD-binary system.


This paper is organized as follows. In Section \ref{section:setup}, we describe our simulation setup and diagnostics. In Section \ref{section:results}, we present the simulation results, especially on the impact of different cooling/heating prescriptions on binary orbital evolution, accretion variability and disc morphology. We also briefly discuss a more comprehensive parameter survey of different temperature and viscosity profiles. Finally, we conclude in Section \ref{section:summary}.

\section{Problem Setup}
\label{section:setup}

In this study, we use the grid-based Godunov code \texttt{ATHENA++} \citep{2020ApJS..249....4S} to solve the vertically integrated viscous hydrodynamic equations in 2D cartesian coordinates $(x,y)$ (similar to \citealt{2019ApJ...875...66M}). The continuity, momentum and energy equations in conservative form read
\begin{align} 
    & \frac{\partial \Sigma}{\partial t} + \nabla \cdot (\Sigma {\bm v}) = s_{\Sigma} , \\ 
    & \frac{\partial (\Sigma \bm{ v})}{\partial t} + \nabla \cdot (\Sigma {\bm v}{\bm v} + P \mathcal{I} )=  s_{{\bm p}} -\Sigma \nabla \Phi - \nabla \cdot {\mathcal T}_{\rm{visc}}, \\
    & \frac{\partial E}{\partial t} + \nabla\cdot[(E+P){\bm v}] = s_{E} -\Sigma {\bm v}\cdot\nabla\Phi + \Lambda - \nabla\cdot ({\mathcal T}_{\rm{visc}}{\bm v}),
\end{align}
where $\Sigma$, ${\bm v}$ are the disc surface density and velocity, $P$ is the vertically integrated pressure, $\mathcal{I}$ is the identity tensor, $\Phi$ is the gravitational potential, ${\mathcal T}_{\text{visc}}$ is the viscous stress tensor, $E$ is the total energy density, $\Lambda$ is the cooling term (to be specified in Section \ref{section:cooling}), and $s_{\Sigma}$, $s_{{\bm p}}$ and $s_{E}$ are the sink terms for surface density, momentum and energy~(to be specified in Section \ref{subsection:sink}). The viscous stress tensor is given in the form of
\begin{equation}
    \mathcal{T}_{\mathrm{visc}, i j}=-\Sigma \nu\left(\frac{\partial v_{i}}{\partial x_{j}}+\frac{\partial v_{j}}{\partial x_{i}}-\frac{2}{3} \frac{\partial v_{k}}{\partial x_{k}} \delta_{i j}\right) ,
\end{equation}
with $\nu$ being the kinematic viscosity, which will be discussed in detail in Section \ref{subsection:tempandvisc-main}. The total energy density is given by
\begin{equation}
    E=\frac{1}{2}\Sigma v^2+\frac{P}{\gamma-1}\ ,
\end{equation}
where $\gamma=5/3$ is the adiabatic index. \st{Viscous heating is automatically included in the code, which enters the energy equation in the form of} \im{The viscous heating term reads}   
\begin{equation}
    Q_{\rm{visc}} = -\nabla \cdot \left(\mathcal{T}_{\rm{visc}}\bm{v}\right) ,
\label{eq:vischeating}
\end{equation}
\im{which directly enters the energy equation.}

Same as \citet{paper0}, we assume equal-mass binary on a fixed circular orbit,
with the mass of \dong{each component} $M_1=M_2=0.5M_{\rm{B}}$ and \dong{the binary} separation $a_{\rm{B}}$. Here we denote $M_{\rm{B}}$ as the total mass of the binary. The gravitational potential of the binary can be expressed as 
\begin{equation}
\begin{aligned} 
\Phi({\bm{r}},t) &= \Phi_1 + \Phi_2 
\label{eq:gravpot} \\
&= - \sum_{i=1}^{2} \frac{GM_i}{\left(|{\bm r} - {\bm r}_i|^2+\epsilon^2\right)^{1/2}}, 
\end{aligned}
\end{equation}
with the gravitational softening length $\epsilon=0.025a_{\rm{B}}$, \dong{where ${\bm r}$ specifies the location of gas element and ${\bm r}_i$ the location of each binary component.} The binary rotates counter-clockwise with Keplerian speed and the CBD evolves under this rotating potential.

Our coordinate system centers on the center-of-mass of the binary with computational domain covering $[-20a_{\rm{B}},20a_{\rm{B}}]\times[-20a_{\rm{B}},20a_{\rm{B}}]$ in $x, y$ direction respectively. At the root level, the resolution is $256\times256$. \haiyang{We take the advantage of adaptive mesh refinement (AMR) in \texttt{Athena++} and apply 4 more levels of refinement. In the domain covering the CBD cavity ($r \leq 2.5 a_{\rm{B}}$), the resolution is the highest, with $\Delta x = \Delta y = 0.0098 a_{\rm{B}}$. We also ensure that at least one (two) more refinement level is applied within $[-10a_{\rm{B}},10a_{\rm{B}}]\times[-10a_{\rm{B}},10a_{\rm{B}}]$ 
 ($[-5a_{\rm{B}},5a_{\rm{B}}]\times[-5a_{\rm{B}},5a_{\rm{B}}]$) comparing to the root grid. }
The accretion process is simulated through sink cells in the vicinity of the binary (see Section \ref{subsection:sink} for details). We roughly resolve the sink cell with 10 grid cells along either the horizontal or vertical direction. At the boundary in both $x$ and $y$ directions, we use the outflow boundary condition. We also impose a damping zone between $[r_{\rm{in}},r_{\rm{out}}] = [18a_{\rm{B}}, 20a_{\rm{B}}]$ (e.g., \citealp{2019ApJ...875...66M}) to quench modes characterized by azimuthal number $m=4$, \dong{to suppress} the impact of the square boundary, and to maintain the constant mass supply: 
\begin{equation}
    \frac{d u}{d t} = -\frac{(r-r_{\rm{in}})r}{(r_{\rm{out}}-r_{\rm{in}})r_{\rm{out}}} \frac{(u-u_{\rm{eq}})}{t_{\rm{damp}}} ,
\end{equation}
with $u$ representing hydrodynamic variables: $(\Sigma, P, v_r, v_{\phi})$, $u_{\rm{eq}}$ being the initial value of these variables, $t_{\rm{damp}}=10^{-3}/\Omega_{\rm{B}}$ and $\Omega_{\rm{B}}=\sqrt{GM_{\rm{B}}/{a_{\rm{B}}^3}}$ the orbital frequency of the binary. 


\begin{figure*}
\begin{center}
\includegraphics[width=0.85\textwidth,trim={0cm 0cm 0cm .0cm},clip]{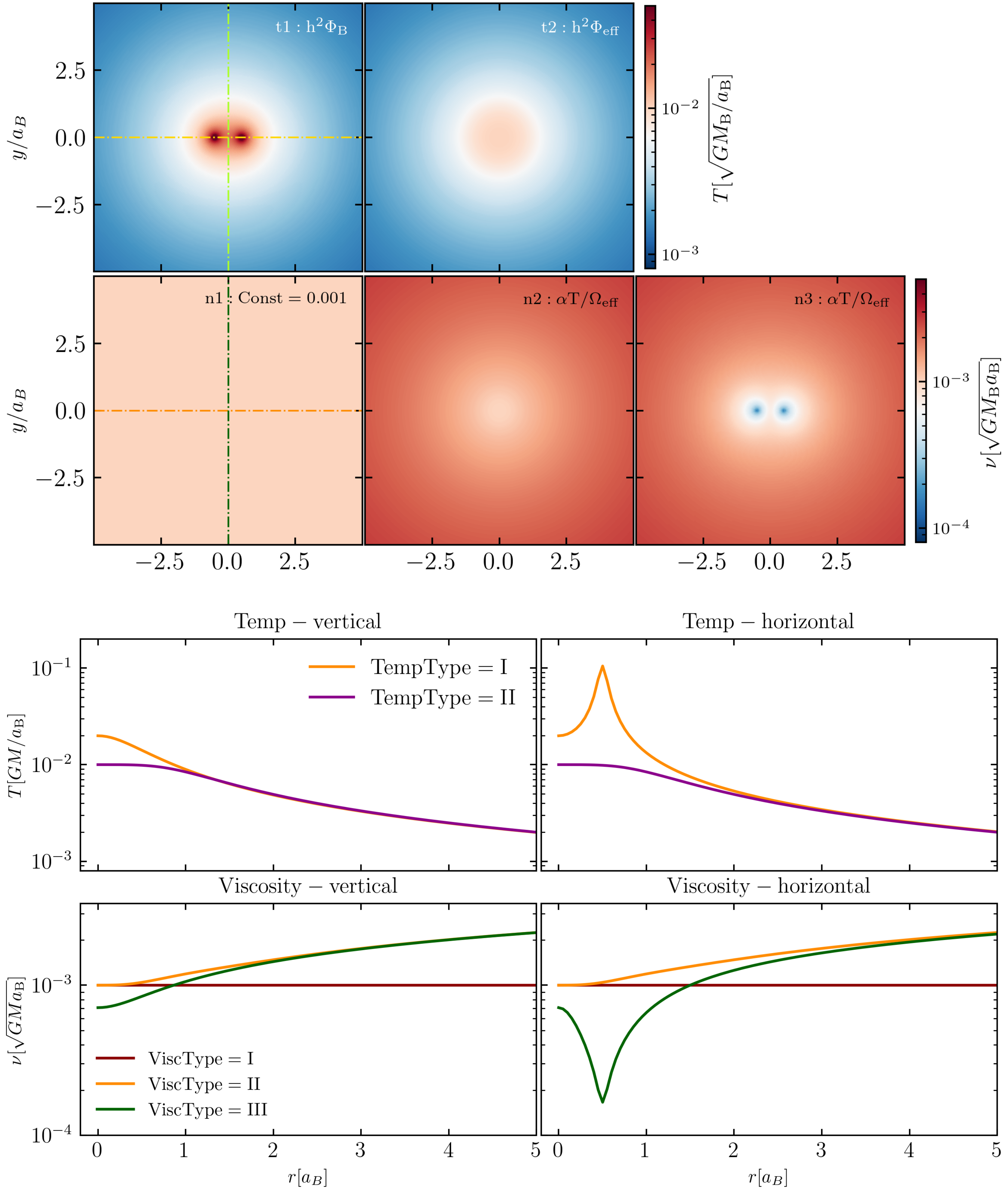}
\end{center}
\vspace{-0.3cm}
\caption{2D and 1D illustrations of the ``equilibrium'' temperature profiles and the viscosity profiles employed in our simulations. The 2D temperature and viscosity distributions are shown in the upper panels. The first and the second rows in the lower panels show the 1D temperature and viscosity profiles along the green dash-dotted vertical and yellow dashed horizontal lines in the 2D plots. \label{fig:tempnu}}
\vspace{-0.3cm}
\end{figure*}

\subsection{``Equilibrium'' Temperature Profile and Viscosity Prescription}
\label{subsection:tempandvisc-main}

\dong{Previous simulations of CBDs have adopted either locally or globally isothermal equation of state. In this work, we solve the energy equation of the flow and include gas cooling using a thermal relaxation ansatz (see Section \ref{section:cooling}). To this end, we need to specify the gas ``equilibrium'' temperature profile.}
which is the target of the thermal relaxation process.



\subsubsection{``Equilibrium'' Temperature Profile}
\dong{We consider} two types of ``equilibrium'' temperature profiles in this paper:

(i) ${\rm Temp \ Type\ I \ (t1):}$ 
\begin{equation}
T_{\rm{eq}} \equiv c_s^2 = h^2 |\Phi| ,
\label{eq:tempprofile1}
\end{equation}
where $c_s$ is the sound speed and $\Phi$ is \dong{given by} Equation \ref{eq:gravpot}. \dong{Note that when} close to one of the binary components, $|{\bm r}-{\bm r}_i|\ll|{\bm r}-{\bm r}_j|$, \dong{we have} $T_{\rm{eq}} \rightarrow h^2 GM_{i}/|{\bm r}-{\bm r}_i|$; and when far from the binary, $|{\bm r}|\gg|{\bm r}_1|, |{\bm r}_2|$, \dong{we have} $T_{\rm{eq}}  \rightarrow h^2 GM_{\rm{B}}/|{\bm r}|$.

(ii) ${\rm Temp \ Type\ II \ (t2):}$ 
\begin{align}
T_{\rm{eq}} &\equiv c_s^2 = h^2 \Phi_{\rm eff} 
\label{eq:tempprofile2} \\
				&= h^2 (|\Phi_{\rm{B}}|^{-n} + |\Phi_K|^{-n})^{-1/n} ,  \nonumber 
\end{align}
where $\Phi_K=-GM_{\rm{B}}/r$ and $\Phi_{\rm{B}}=-GM_{\rm{B}}/a_{\rm{B}}$, which acts as a ``softening'' potential to avoid singularity in temperature profile\dong{; we adopt $n=4$. Obviously, the two temperature profiles are the same for $r \gg a_{\rm{B}}$, but can differ significantly for $r \lesssim a_{\rm{B}}$}.

\subsubsection{Viscosity Prescription}
In most CBD studies, viscosity adopted in the simulations is either constant or $\alpha-$viscosity \citep{1973A&A....24..337S}. 
When adopting the $\alpha-$viscosity, the viscosity profile $\nu = \alpha c_s H$ ($H$ is the disc scale height) can be centered around each of the binary component or the coordinate center, similar to the two different  temperature profile types.
But in some cases, employing such an $\alpha-$viscosity in the simulation can be problematic: The viscous time in the disc scale grows quadratically with the Mach number $v_{\rm{K}}/c_s$, making \dong{it computationally challenging for} thin discs ($h\lesssim0.05$) to reach a quasi-steady state. Consequently, constant kinematic viscosity coefficients are adopted by some studies (e.g., \citealt{2020ApJ...900...43T,2020ApJ...901...25D,2021ApJ...921...71D,2022MNRAS.513.6158D}). \dong{In this paper, we adopt two viscosity prescriptions mentioned above}. In order to straightforwardly compare simulations adopting different viscosity profiles, all of the simulations in this study \dong{use} $\alpha=0.1$ and $h=0.1$.

(i) ${\rm Visc \ Type\ I \ (n1):}$ When adopting constant kinematic viscosity, we \dong{choose}
\begin{equation}
\nu = \alpha h^2 a_{\rm{B}}^2 \Omega_{\rm{B}} ,
\label{eq:viscprofile1}
\end{equation}
with $\alpha=0.1$ and $h=0.1$, \dong{which gives} $\nu = 10^{-3} a_{\rm{B}}^2 \Omega_{\rm{B}}$.

(ii) ${\rm Visc \ Type\ II \ (n2):}$ When adopting $\alpha-$viscosity, we can have two different types. The viscosity profile depends on the temperature profile chosen in the simulation. In the simulations employing an axisymmetric equilibium temperature profile \dong{(see Equation \ref{eq:tempprofile2})}, the viscosity profile is also axisymmetric
\begin{equation}
\begin{aligned}
\nu &= \alpha c_s H = \alpha \frac{T}{\Omega_{\rm eff1}} \\
				&=  \alpha \frac{T}{ 
 (|\Omega_{\rm{B}}|^{-n} + |\Omega_K|^{-n})^{-1/n} } ,
\label{eq:viscprofile2}
\end{aligned}
\end{equation}
where we choose $n=2$.

(iii) ${\rm Visc \ Type\ III \ (n3):}$ In the simulations employing the temperature profile which centers around each of the binary component, the viscosity profile is 
\begin{equation}
\begin{aligned}
\nu &= \alpha c_s H = \alpha \frac{T}{\Omega_{\rm eff2}} \label{eq:viscprofile3} \\
&=  \alpha \frac{T}{ 
 (|\Omega_1|^{-n} + |\Omega_2|^{-n})^{-1/n} } , 
\end{aligned}
\end{equation}
where $n=2$, $\Omega_1=\sqrt{GM_1/(|{\bm r}-{\bm r}_1|^2+\epsilon^2)^{3/2}}$, and $\Omega_2=\sqrt{GM_2/(|{\bm r}-{\bm r}_2|^2+\epsilon^2)^{3/2}}$. Here we choose $n=2$ so that the viscosity profile is similar to the one widely used in previous studies $\nu = \alpha T/{\Omega_{\rm{eff}}}$ (e.g., \citealt{2016ApJ...827...43M,2019ApJ...871...84M}). Same as \citet{2019ApJ...871...84M}, $\Omega_{\rm{eff}}$ reduces to $\sqrt{GM_{\rm{B}}/|{\bm r}|}$ far from the binary and to $\sqrt{GM_{\rm{i}}/|{\bm r}-{\bm r}_i|^{3}}$ close to one of the binary components. Consequently, the viscosity reduces to 
\begin{equation}
    \nu \rightarrow \alpha \frac{T}{\Omega_K} = \alpha \frac{GM_{\rm{B}}/|{\bm r}|}{\sqrt{GM_{\rm{B}}/|{\bm r}|^3}} ,
\end{equation}
when $|{\bm r}|\gg|{\bm r}_1|, |{\bm r}_2|$, which is equivalent to the $\alpha-$viscosity in accretion discs hosting single binary component. And when $|{\bm r}-{\bm r}_i|\ll|{\bm r}-{\bm r}_j|$, the viscosity reduces to 
\begin{equation}
    \nu \rightarrow \alpha \frac{T}{\Omega_i} = \alpha \frac{GM_{\rm{i}}/|{\bm r}-{\bm r_i}|}{\sqrt{GM_{\rm{i}}/|{\bm r}-{\bm r_i}|^3}} ,
\end{equation}
In this way, the viscosity in each of the mini-disc can also be regarded as standard $\alpha-$viscosity. The 2D/1D temperature and viscosity profiles are separately illustrated in the upper/lower panels of Figure \ref{fig:tempnu}. 

\subsection{Thermal Relaxation}
\label{section:cooling}

As in \citet{paper0}, we thermally relax gas temperature to the prescribed \dong{``equilibrium''}value of $T_{\rm eq}$ over a cooling timescale $t_{\rm cool}$, controlled by the cooling term
\begin{equation} 
\Lambda=-\frac{\Sigma}{\gamma - 1} \times \frac{(T - T_{\text{eq}})}{t_{\text{cool}}}\ ,
\label{eq:cooling}
\end{equation}
We prescribe the cooling time in a scale-free manner, using the dimensionless parameter
\begin{equation}
    \beta\equiv\Omega_{\rm{eff2}}t_{\rm cool}\ ,
\end{equation}
known as dynamical cooling (or $\beta-$cooling)~\citep{2001ApJ...553..174G}. \dong{Here the} effective ``orbital frequency'' \dong{is given by (see Equation \ref{eq:viscprofile3})}.
\begin{equation}
    \Omega_{\rm{eff2}} = \left( |\Omega_1|^{-n} + |\Omega_2|^{-n} \right)^{-1/n}
\end{equation} 

The temperature of the CBD is expected to reach a new equilibrium $T_{\rm{eq}}^{\prime}$ as a result of the combined effects of dynamical cooling and viscous heating \im{(see, e.g., \citealt{2022A&A...664A.157S})}. 
When assuming $\alpha-$viscosity, the latter is $Q_{\text {visc }} \approx \frac{9}{4} v \Sigma \Omega_{\mathrm{K}}^2$ (e.g., \citealt{1985apa..book.....F}). \dong{Neglecting} the effect of compressional heating $(\nabla \cdot \boldsymbol{u}=0)$, we can expect a modified temperature profile $T_{\rm{eq}}^{\prime}$
\begin{align}
	&Q_{\text {visc }}+\Lambda=0 , \\
	\Rightarrow &T_{\rm{eq}}^{\prime}=\frac{T_{\rm{eq}}}{1-k \alpha \beta}; \quad k=\frac{9}{4} (\gamma-1).
\label{eq:eqtemp}
\end{align}
As the viscosity and cooling time increases, the new equilibrium temperature will increase accordingly. 

To determine the range of cooling parameter $\beta$ in our survey, we should refer to realistic scenarios, as the temperature distribution in CBDs is determined by the balance of radiative cooling and viscous heating.
Following Equation \ref{eq:eqtemp} and requiring the denominator to be non-zero, \dong{we find} $\beta \lesssim 6$. \dong{In the following, we will} conduct simulations with $\beta \in [0.0, 0.2, 0.5, 1.0, 2.0, 4.0]$.

\im{It is worth noticing that adopting constant-$\beta$ dynamical cooling is a simplified version of radiative cooling. 
In thin (and optically thick) disk limit, the effective beta parameter for radiative cooling depends on the specific parameters of the system, including binary mass, accretion rates, spatial location, as well as opacity. With such complications, the simulations cannot be scale-free. As a first exploration to demonstrate the role of thermodynamics, the beta cooling prescription is much more handy and we thus refrain from explicitly considering radiative cooling treatment. We refer to \citealt{2022A&A...664A.157S} as an example for radiative cooling CBDs around young stellar objects (e.g., see their Fig. 7).}




\subsection{Initial Condition}

We use axisymmetric initial conditions with the outer CBD in an approximately viscous steady state. Focusing on fixed circular orbit, equal-mass binary with disc aspect ratio $h=0.1$, the surface density profile in the initial setup is given by
\begin{equation}
	\Sigma(r) = \Sigma_0(r) \rm{exp} \left[ - \left( \frac{r}{r_{\rm{edge}}}\right)^{-6} \right] ,
\label{eq:surfden}
\end{equation}
with $r_{\rm{edge}} = 2.5a_{\rm{B}}$ and $\Sigma_0(r) = \dot M_0/(3\pi\nu)$. The exponential cutoff factor creates an artificial cavity around the binary and $r_{\rm{edge}}$ is an initial guess of the cavity radial size. Here $\nu$ depends on our choice of viscosity prescription.

The initial radial velocity is set to be 
\begin{equation}
	v_r(r) = -\frac{3 \nu}{2 r}  \rm{exp} \left[ - \left( \frac{r}{r_{\rm{edge}}}\right)^{-6} \right] ,
\label{eq:radvel}
\end{equation} 
And we choose initial angular velocity to be 
\begin{equation}
	\Omega(r) = \left[\Omega_0(r)^{-4}+\Omega_{\rm{B}}^{-4}\right]^{-1/4} ,
\end{equation}
where $\Omega_0$ is the equilibrium angular velocity far from the binary
\begin{equation}
	\Omega_0 = \sqrt{ \frac{GM_{\rm{B}}}{r^3}(1-h^2) } ,
\label{eq:azivel}
\end{equation}
We apply an exponential cutoff in the initial radial velocity profile in Equation \ref{eq:radvel} and a ``softened'' initial azimuthal velocity profile in Equation \ref{eq:azivel} inside the CBD cavity to avoid violent relaxation at the beginning of the simulation. Our simulations are scale-free, with $G=M_{\rm{B}}=a_{\rm{B}}=\Omega_{\rm{B}}=1$ in code units, thus the binary period is $P_{\rm{B}}=2\pi$. As we have stated in Section \ref{subsection:tempandvisc-main}, viscosity is parameterized by choosing $\alpha=0.1$ and $h=0.1$. And the mass injection rate is given as $\dot M_0 = 3 \pi \alpha h^2$. 

\subsection{The Sink Term}
\label{subsection:sink}

As the accretion of each \dong{binary component} is simulated through the sink cells, we remove the mass, angular momentum, and energy according to the torque-free scheme (similar to, e.g., \citealt{2020ApJ...892L..29D,2022MNRAS.513.6158D}). The sink terms for surface density $s_{\Sigma}$, momentum $s_{\bm{p}}$, and energy $s_{E}$ are given as follows:
\begin{align}
	s_{\Sigma} &= s_{\Sigma,1}+s_{\Sigma,2} \nonumber \\
        &= -\frac{\Sigma}{t_{s}} \left[ {\rm exp}\left( -\left( \frac{|{\bm r}-{\bm r}_1|}{r_s} \right)^{4} \right) +  {\rm exp}\left( -\left( \frac{|{\bm r}-{\bm r}_2|}{r_s} \right)^{4} \right) \right] , \\
	s_{\bm{p}} &= s_{\bm{p},1}+s_{\bm{p},2} \nonumber \\
        &= -\frac{\Sigma}{t_{s}} \left[ {\rm exp}\left( -\left( \frac{|{\bm r}-{\bm r}_1|}{r_s} \right)^{4} \right) {\bm v}^{\rm{res}}_1 +  {\rm exp}\left( -\left( \frac{|{\bm r}-{\bm r}_2|}{r_s} \right)^{4} \right){\bm v}^{\rm{res}}_2 \right]  , \\
        s_{E} &= s_{E,1}+s_{E,2} \nonumber \\
        &= -\frac{\Sigma}{t_{s}} \left[ {\rm exp}\left( -\left( \frac{|{\bm r}-{\bm r}_1|}{r_s} \right)^{4} \right) E^{\rm{res}}_1 +  {\rm exp}\left( -\left( \frac{|{\bm r}-{\bm r}_2|}{r_s} \right)^{4} \right) E^{\rm{res}}_2 \right]  , \\
    {\bm v}_i^{\rm{res}} &= ({\bm v}-{\bm v}_i)\cdot \hat r_i \hat r_i+{\bm v}_i , \\
    {E}_i^{\rm{res}} &= \frac{1}{2} {\bm v}_i^{\rm{res}}\cdot{\bm v}_i^{\rm{res}} + \frac{T}{\gamma-1} ,
    \label{eq:velres}
\end{align}
where $s_{\Sigma,i}$, $s_{\bm{p},i}$, and $s_{E,i}$ are separately the sink terms for surface density, momentum and energy of each binary component, $r_s = 0.05a_{\rm{B}}$ is the radius of the sink cell, $t_{\rm{s}}$ is the sink timescale; ${\bm v}_i^{\rm{res}}$ is the ``extracted'' velocity of the sink cell in comoving frame of the binary, ${\bm v_i}$ is the velocity of the $i\rm{th}$ binary component, ${E}_i^{\rm{res}}$ is the ``extracted'' specific energy of the sink cell, and $\hat r_i$ is the unit basis vectors in the polar coordinate centered on each binary component in radial direction. 

In the comoving frame of the sink cell, only momentum in the radial direction is absorbed. \im{The major difference between the standard sink term and torque-free sink term is that the latter only extracts orbital angular momentum (i.e., angular momentum relative to the center of mass of the binary) and does not extract spin angular momentum (i.e., angular momentum relative to each accreting binary component) associated with the mass accreted in the sink radius.} \im{Namely, the torque-free sink does not exert a torque on the fluid in the frame of a single binary component.}
For most of the simulations, we choose sink time $t_{s}=(\gamma_{\rm{s}}\Omega_{\rm{B}})^{-1}$ with $\gamma_{\rm{s}}=3.0$. The only exception is that in the simulations adopting dynamically varying $\alpha-$viscosity and axisymmetric temperature profile (namely, runs \texttt{t2n2b-} in Table \ref{tab:simulation-parameters-maintext}), we instead choose the sink time to be 
 \begin{equation}
     t_{s}=\frac{2|{\bm r}-{\bm r}_i|^2}{3\nu} .
\label{eq:tsink}
 \end{equation}
\dong{Choosing this $t_s$ would avoid mass accumulation, and the final results are barely affected by this different sink prescription as we will show in Section }\ref{subsection:survey}.

\subsection{Diagnostics}
\label{subsection:diagnostics}

All of the simulations are run for 2200 binary orbits. After reaching a quasi-steady state, the net rate of angular momentum transfer across the CBD should be the same as the angular momentum transfer rate between CBD and the binary. In this work, we separately calculate these transfer rates. First, the total torque $\dot J_{\rm{tot}}$ \dong{on the binary can be computated} as the sum of accretion torque $\dot J_{\rm{acc}}$ and gravitational torque $\dot J_{\rm{grav}}$, \dong{given by}
\begin{align}
    &\dot J_{{\rm tot}} = \dot J_{\rm{acc}} + \dot J_{\rm{grav}} , \\
    &\dot J_{{\rm acc},i}|_{i=1,2} = \boldsymbol{\hat e_z} \cdot\int {\bm r} \times s_{\bm{p},i} dA , \\
    &\dot J_{{\rm grav},i}|_{i=1,2} = -\boldsymbol{\hat e_z} \cdot\int {\bm r} \times (\Sigma\nabla\Phi_i) dA ,
\end{align}
Because binary orbital evolution is determined by the specific torque ($\dot J_{\rm{tot}}/\dot M$, where $\dot M = \dot M_1+\dot M_2$), we also store the accretion rate of each sink cell 
\begin{equation}
    \dot M_i|_{i=1,2} = -\int s_{\Sigma,i} dA ,
\end{equation}
These values are outputted as time series every $1/100$ binary orbit. We use the time-averaged specific torque in the last 200 orbits to determine the orbital evolution of the binary. 

The second method is to calculate the angular momentum flux in the CBD.
Same as \citet{paper0}, the advective, viscous, gravitational and total torques along with the accretion rate are given by (e.g., \citealt{2017MNRAS.466.1170M,2019ApJ...871...84M,2019ApJ...875...66M}):
\begin{flalign}
{{\dot J}}_{\mathrm{adv}} &= \boldsymbol{\hat e_z} \cdot \oint \Sigma \boldsymbol{v} \times \boldsymbol{r}(\boldsymbol{v} \cdot \boldsymbol{r}) d\phi \\
&= -  \oint r^2 \Sigma v_r v_{\phi} d\phi \nonumber , \\
 \vspace{-0.5cm}
{{\dot J}}_{\mathrm{visc}} &= -\boldsymbol{\hat e_z} \cdot \oint\left(\mathcal{T}_{\mathrm{visc}} \cdot {\boldsymbol{r}}\right)  \times \boldsymbol{r} d\phi , \\
&= -  \oint r^3 \nu \Sigma \left[\frac{\partial}{\partial r}\left(\frac{v_{\phi}}{r}\right) + \frac{1}{r^2}\frac{\partial v_r}{\partial \phi}\right] d\phi , \nonumber \\ 
\vspace{-0.5cm}
{T}_{\mathrm{grav}} &= \boldsymbol{\hat e_z} \cdot \int_{r}^{r_{\rm{out}}} \left[- \oint \Sigma(\boldsymbol{r} \times \boldsymbol{\nabla} \Phi) d\phi \right] dr \\
&= \int_r^{r_{\mathrm{out}}} \left(- \oint r \Sigma \frac{\partial \Phi}{\partial \phi} d\phi\right) dr \nonumber , 
 \vspace{-1.0cm}
\end{flalign}
\begin{flalign}
&{{\dot J}}_{\mathrm{tot}} = {{\dot J}}_{\mathrm{adv}}-{{\dot J}}_{\mathrm{visc}}-{T}_{\mathrm{grav}} , \\
&\dot{M} 
= -\oint \Sigma v_r r d\phi , 
\end{flalign}
These radial profiles of angular momentum flux are calculated in each timestep, and time-averaged results are generated by post-processing the outputted, time-integrated values. Same as the first method, we use time-averaged results from the last 200 binary orbits when the inner disc has reached a quasi-steady state. 

For equal mass, circular binary, \dong{the secular orbital evolution rate is given by}
\begin{equation}
    \frac{\dot a}{a_{\rm{B}}} = 8\left( \frac{l_0}{\Omega_{\rm{B}}a_{\rm{B}}^2} - \frac{3}{8}\right) \frac{\dot M}{M_{\rm{B}}} ,
\end{equation}
where the total torque per unit of accreted mass is
\begin{equation}
    l_0=\langle\dot J_{\rm{tot}}\rangle/\langle\dot M\rangle\ .
\end{equation}
Therefore, whether binary shrink or expand depends on whether $l_0$ is greater or smaller than $3\Omega_{\rm{B}}a_{\rm{B}}^2/8$.

To quantify the evolution of the disc morphology, we also calculate the mass-weighted eccentricity of the cavity
\begin{align}
    \langle e \rangle_x &= \frac{\int^{6a_{\rm{B}}}_{a_{\rm{B}}} \int^{2\pi}_0 e_x \Sigma rdrd\phi}{\int^{6a_{\rm{B}}}_{a_{\rm{B}}} \int^{2\pi}_0 \Sigma rdrd\phi} , \\
    \langle e \rangle_y &= \frac{\int^{6a_{\rm{B}}}_{a_{\rm{B}}} \int^{2\pi}_0 e_y \Sigma rdrd\phi}{\int^{6a_{\rm{B}}}_{a_{\rm{B}}} \int^{2\pi}_0 \Sigma rdrd\phi} ,
\end{align}
where the eccentricity vector and its the $x, y$ component are 
\begin{align}
    {\bm e} &=\frac{{\bm v}^2 {\bm r}-({\bm v} \cdot {\bm r}) {\bm v}}{G M_{\rm{B}}}-\hat{{\bm r}} , \\
    e_x &=\frac{r v_r v_\phi}{G M_{\rm{B}}} \sin \phi+\left(\frac{r v_\phi^2}{G M_{\rm{B}}}-1\right) \cos \phi , \\
    e_y &=-\frac{r v_r v_\phi}{G M_{\rm{B}}} \cos \phi+\left(\frac{r v_\phi^2}{G M_{\rm{B}}}-1\right) \sin \phi .
\end{align}
Similar to mass-weighted eccentricity, we calculate the mass dipole of the entire CBD. 
\begin{equation}
   \Psi= \int^{18a_{\rm{B}}}_{0} \Sigma {\rm e}^{i \phi} r d r d \phi ,
\end{equation}
where the upper limit of the integration is chosen to be the inner radius of the damping zone. \dong{The quantity monitors} the growth and damping of any asymmetric instability in the CBD, including disc eccentricity and spiral features. 
We also output 2D snapshots of various fluid quantities (such as surface density $\Sigma$, temperature $T$, viscosity $\nu$, and the
gravitational torque density $T^{2D}_{\rm{grav}}=-\Sigma {\bm r}\times \nabla \Phi$ every 10 orbits. 

\begin{table*}
	\centering
	\begin{tabular}{cccccccc}
	\hline
	\hline
	ID & $\nu$(Type)  &$T$(Type) &$\beta$ & $\langle\dot J_{\text{acc}}\rangle/\langle\dot M\rangle$  & $\langle\dot J_{\text{grav}}\rangle/\langle\dot M\rangle$ & $\langle\dot J_{\text{tot}}\rangle/\langle\dot M\rangle$\\
	\hline
        Fiducial Runs &&&&&& \\
        \hline
        \texttt{t1n1b0.0} & Const (I) & I & 0.0  & 0.251 &  0.458 &  0.709 \\
        \texttt{t1n1b0.2} &           &   & 0.2  & 0.251 &  0.100 &  0.351 \\
        \texttt{t1n1b0.5} &           &   & 0.5  & 0.250 & -0.079 &  0.172 \\
        \texttt{t1n1b1.0} &           &   & 1.0  & 0.249 & -0.078 &  0.172 \\
        \texttt{t1n1b2.0} &           &   & 2.0  & 0.248 &  0.033 &  0.281 \\
        \texttt{t1n1b4.0} &           &   & 4.0  & 0.248 & -0.738 & -0.491 \\   
        \hline   
        \texttt{t1n3b0.0} & $\alpha-$visc (III) & I & 0.0  & 0.253 &  0.506 &  0.759 \\
        \texttt{t1n3b0.2} &                     &   & 0.2  & 0.253 &  0.363 &  0.616 \\
        \texttt{t1n3b0.5} &                     &   & 0.5  & 0.253 & -1.246 & -0.993 \\
        \texttt{t1n3b1.0} &                     &   & 1.0  & 0.253 &  0.292 &  0.545 \\
        \texttt{t1n3b2.0} &                     &   & 2.0  & 0.252 &  0.535 &  0.786 \\
        \texttt{t1n3b4.0} &                     &   & 4.0  & 0.250 &  0.439 &  0.689 \\
	\hline
        Parameter Survey &&&&&& \\
	\hline
        \texttt{t2n1b0.0} & Const (I) & II & 0.0 & 0.250 & 0.429  & 0.679 \\
        \texttt{t2n1b0.2} &       &    & 0.2     & 0.250 & 0.093  & 0.344 \\
        \texttt{t2n1b0.5} &       &    & 0.5     & 0.250 & -0.078 & 0.173 \\
        \texttt{t2n1b1.0} &       &    & 1.0     & 0.250 & -0.161 & 0.089 \\
        \texttt{t2n1b2.0} &       &    & 2.0     & 0.250 & -0.086 & 0.164 \\
        \texttt{t2n1b4.0} &       &    & 4.0     & 0.250 & -0.525 & -0.276 \\
	\hline
        \texttt{t2n2b0.0} & $\alpha-$visc (II) & II & 0.0 & 0.250 &0.505 & 0.756 \\
        \texttt{t2n2b0.2} &                    &    & 0.2 & 0.260 &0.371 & 0.631 \\
        \texttt{t2n2b0.5} &                    &    & 0.5 & 0.271 &-1.989& -1.719 \\
        \texttt{t2n2b1.0} &                    &    & 1.0 & 0.261 &0.074 & 0.335 \\
        \texttt{t2n2b2.0} &                    &    & 2.0 & 0.258 &0.177 & 0.434 \\
        \texttt{t2n2b4.0} &                    &    & 4.0 & 0.259 &0.311 & 0.571 \\
        \hline
        \texttt{t1n3b0.0}* & Fixed $\alpha-$visc (III) & I & 0.0 & 0.251 & 0.471  & 0.722 \\
        \texttt{t1n3b0.2}* &                           &   & 0.2 & 0.251 & 0.278  & 0.529 \\
        \texttt{t1n3b0.5}* &                           &   & 0.5 & 0.251 & 0.178  & 0.429 \\
        \texttt{t1n3b1.0}* &                           &   & 1.0 & 0.251 & 0.119  & 0.371 \\
        \texttt{t1n3b2.0}* &                           &   & 2.0 & 0.251 & 0.153  & 0.404 \\
        \texttt{t1n3b4.0}* &                           &   & 4.0 & 0.251 & -0.167 & 0.084 \\
	\hline
        \texttt{t2n2b0.0}* & Fixed $\alpha-$visc (II) & II & 0.0 & 0.250 & 0.505  & 0.756 \\
        \texttt{t2n2b0.2}* &                          &    & 0.2 & 0.250 & 0.326  & 0.577 \\
        \texttt{t2n2b0.5}* &                          &    & 0.5 & 0.250 & 0.226  & 0.477 \\
        \texttt{t2n2b1.0}* &                          &    & 1.0 & 0.250 & 0.163  & 0.413 \\
        \texttt{t2n2b2.0}* &                          &    & 2.0 & 0.250 & 0.136  & 0.386 \\
        \texttt{t2n2b4.0}* &                          &    & 4.0 & 0.250 & -0.062 & 0.188 \\
        \hline
	\end{tabular}
	\caption{Summary of simulation parameters and key results. All of the simulations included in this paper are presented. We have separated simulations to fiducial runs and parameter survey runs. Fiducial runs include 
 two sets of simulations employing constant viscosity and $\alpha-$viscosity, which are discussed in the Section \ref{section:results}. Parameter survey runs include four sets of simulations surveying the impact of different temperature profiles and viscosity prescriptions. Different from adopting standard $\alpha-$viscosity, we adopt fixed $\alpha-$viscosity in the last two subsets (labeled with $*$). The viscosity profiles in these simulations are set to prescribed values instead of varying with local temperature.
 }
	\label{tab:simulation-parameters-maintext}
\end{table*}

\subsection{Simulation Runs}
\label{subsection:run}

We show in Table \ref{tab:simulation-parameters-maintext} the parameters and various specific torque components of all simulations presented in the main text of this paper. In the seven columns, we separately indicate (i) the ID representing the simulations, (ii) the type of viscosity profile, (iii) the type of \dong{equilibrium} temperature profile, (iv) the cooling parameter $\beta$,  (v) the specific accretion torque $\langle\dot J_{\rm{acc}}\rangle/\langle\dot M\rangle$, (vi) the specific gravitational torque $\langle\dot J_{\rm{grav}}\rangle/\langle\dot M\rangle$ and (vii) the specific total torque $\langle\dot J_{\rm{tot}}\rangle/\langle\dot M\rangle$. The IDs characterizing each run are in the form of \texttt{t-n-b-}, including information of the type of temperature profile, viscosity profile and the value of cooling parameter $\beta$ in the simulations. For convenience, we will refer to a certain subset of simulations using \texttt{t-n-b-}. For example, we will first discuss the simulations employing type I viscosity profile and type I temperature profile in Section \ref{section:results} and we use \texttt{t1n1b-} for short.

We separate our simulations into two sets: the fiducial runs and the parameter survey. \haiyang{The convergence tests of simulation resolution, grid configuration, and sink parameters are investigated thoroughly and will be reported in the follow-up work.}
The fiducial runs consists of two subsets of simulations separately adopting constant viscosity profile \texttt{t1n1b-} and $\alpha-$viscosity profile \texttt{t1n3b-}. Section \ref{section:results} will be mainly based on these simulations.

As one of our goals of this study, we also include 4 additional subsets of simulations in the parameter survey to explore the impact of various temperature and viscosity profiles on binary orbital evolution. In the last two subsets of the simulations, we adopt fixed $\alpha-$viscosity profile. Instead of varying locally as the simulation proceeds, the temperature/sound speed influencing the viscosity is set to the prescribed value based on the initial temperature. \dong{For clarity}, these simulations have ID labeled with * in the end: \texttt{t-n-b-}*.

\section{Simulation Results}
\label{section:results}

We separately discuss the combined effect of dynamical cooling and viscous heating on angular momentum transfer between the disc and the binary, accretion variability, and disc morphology. \dong{We first} show the results \dong{using} constant kinematic viscosity in Section \ref{subsec:constvisc}, and then the results employing dynamically varying $\alpha-$viscosity in Section \ref{subsec:varyvisc}. The major difference between the two choices is whether the viscosity will be affected by local temperature or not (namely, whether the viscosity profile is time-dependent). Through this approach, we can \dong{evaluate} how the dependence of viscosity on temperature affects binary migration rate, morphology, and accretion variability. \dong{And as we will show in Section \ref{subsection:survey}, simulations employing fixed $\alpha-$viscosity behave similarly to those employing constant kinematic viscosity. Here we show the constant viscosity runs first in order to comparing with previous studies (e.g., \citealt{2020ApJ...901...25D,2021ApJ...921...71D}).}

\subsection{Results for Constant Viscosity}
\label{subsec:constvisc}

In this subsection we explore the accretion behaviour of CBDs with dynamical cooling and constant viscosity, along with the angular momentum transfer within the disc and between the disc and the binary. 

\begin{figure}
\begin{center}
\includegraphics[width=0.45\textwidth,trim={0cm 0cm 0cm .0cm},clip]{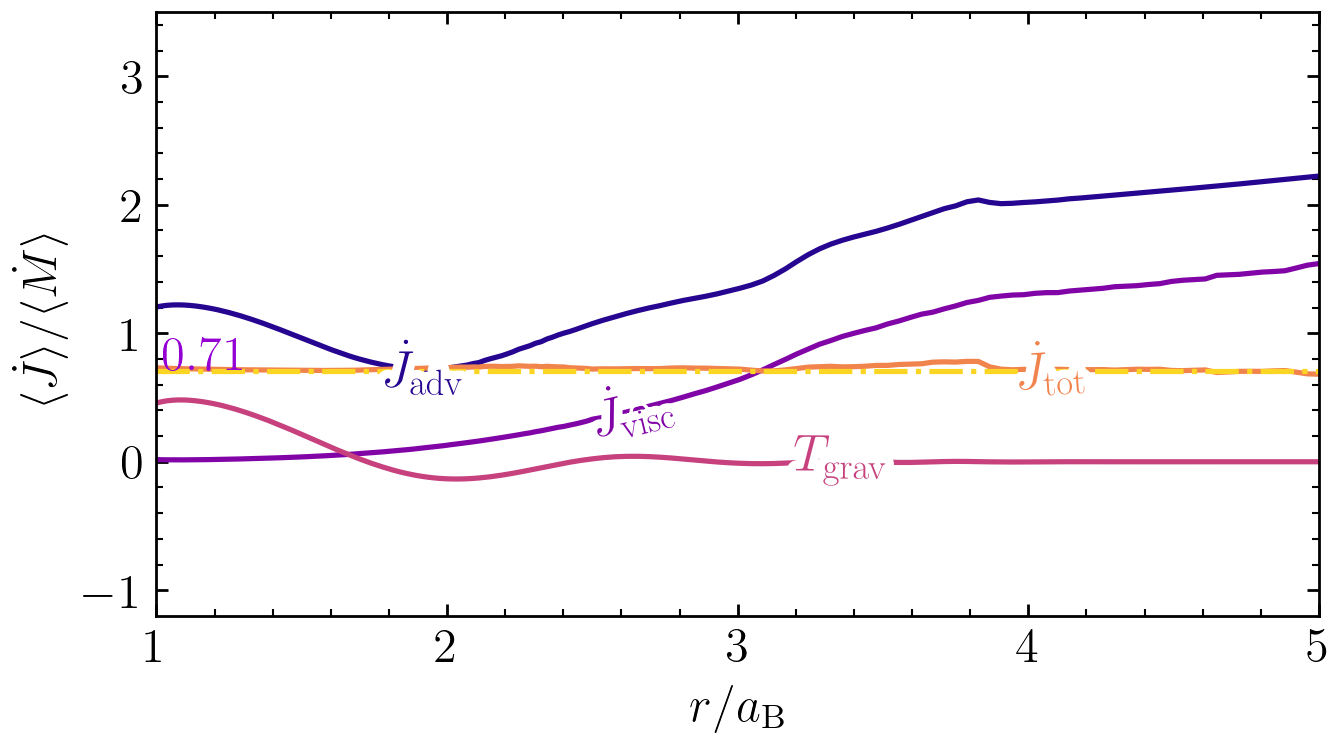}
\end{center}
\vspace{-0.3cm}
\caption{
Radial distribution of the time-averaged advective, viscous, gravitational and total angular momentum current for Run \texttt{t1n1b0.0}, with viscosity $\nu=10^{-3}a^{2}_{\rm{B}}\Omega_{\rm{B}}$ and locally isothermal equation of state. The orange curve is remarkably flat, showing that a quasi-steady state has been reached. For comparison purpose, we overlay a horizontal yellow dashed line with the value $\langle \dot J_{\rm{tot}} \rangle=0.709 \langle \dot M \rangle a_{\rm{B}}^2 \Omega_{\rm{B}}$. 
\label{fig:amcurrent-1}}
\vspace{-0.3cm}
\end{figure}

\begin{figure}
\begin{center}
\includegraphics[width=0.45\textwidth,trim={0cm 0cm 0cm .0cm},clip]{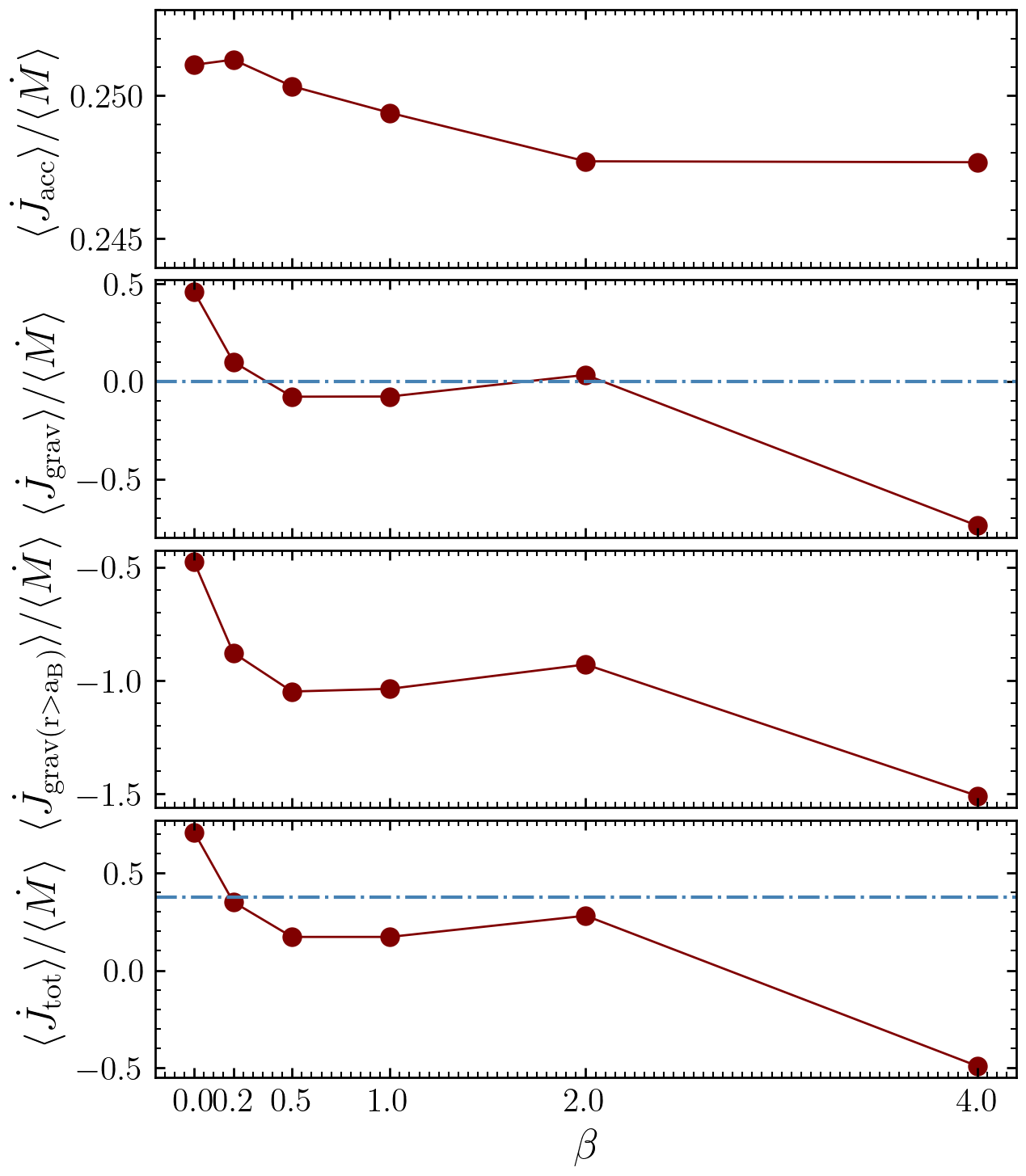}
\end{center}
\vspace{-0.3cm}
\caption{Various specific torque components and the total specific torque of the CBD acting on the binary as a function of parameterized cooling time $\beta$ in fiducial run subset \texttt{t1n1b-}. From the top to the bottom are (1) the specific accretion torque, (2) the specific gravitational torque, (3) the specific gravitational torque from the region $r>a_{\rm{B}}$, and (4) the specific total torque. The dashed blue line in the second panel differentiates the sign of total gravitational torque, and the dashed line in the fourth panel illustrates the threshold of binary orbital evolution.  
\label{fig:torque-subset1}}
\vspace{-0.3cm}
\end{figure}

\begin{figure}
\begin{center}
\includegraphics[width=0.45\textwidth,trim={0cm 0cm 0cm .0cm},clip]{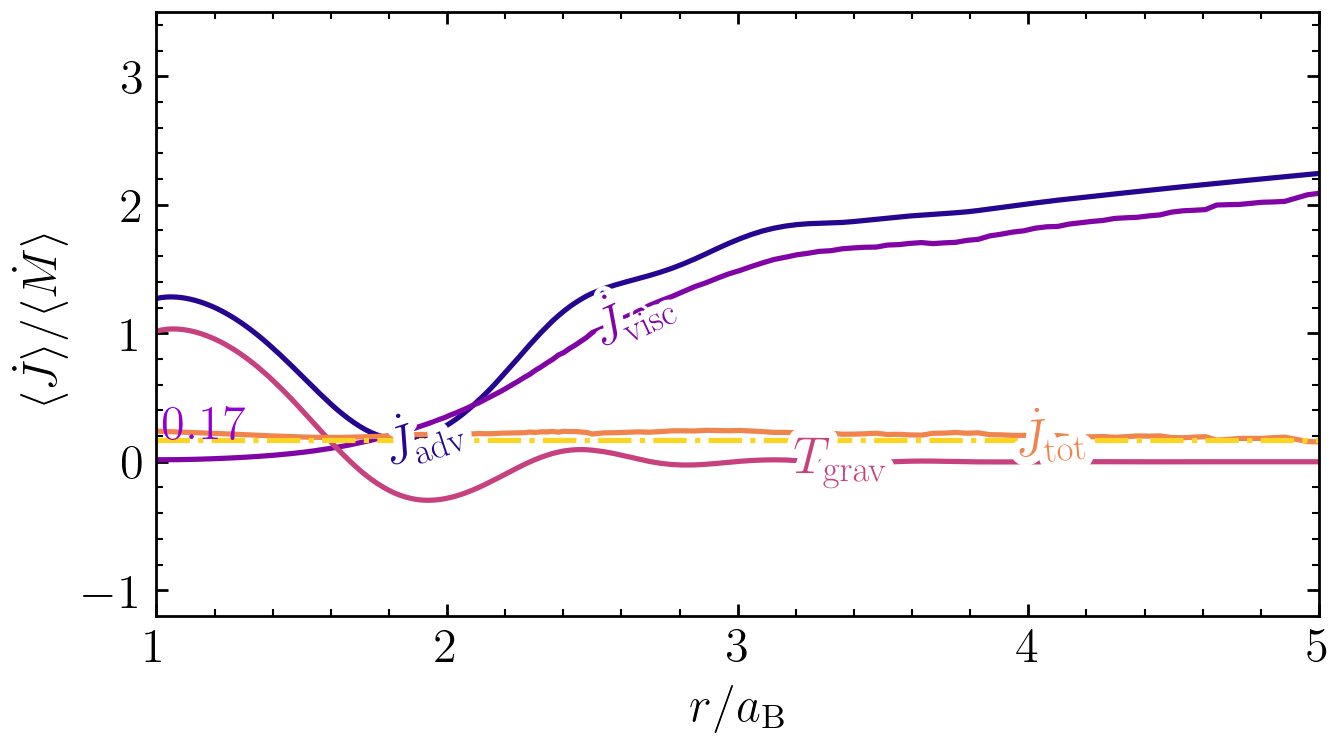}
\end{center}
\vspace{-0.3cm}
\caption{Similar to Figure \ref{fig:amcurrent-1}, except for Run \texttt{t1n1b0.5}, with cooling parameter $\beta=0.5$. \im{The gravitational angular momentum current $T_{\rm{grav}}$ is larger at inner radial location and viscous angular momentum current $\dot J_{\rm{visc}}$ are significantly larger at outer radial location, compared to those in Figure \ref{fig:amcurrent-1}. This results in a significantly smaller total angular momentum current $\dot J_{\rm{tot}}$.}
\label{fig:amcurrent-2}}
\vspace{-0.3cm}
\end{figure}

\subsubsection{Angular Momentum Transfer}
\label{subsubsec:constvisc-amtransfer}

In the locally isothermal limit \dong{$\beta=0$}, the specific torque calculated from our simulation is in close agreement with the results obtained in previous works. The specific accretion torque and the gravitational torque are outputted as a time series. These quantities are then time-averaged in the interval of $[2000P_{\rm{B}}, 2200P_{\rm{B}}]$, after the disc has reached a quasi-steady state. \dong{We} find the specific torque acting on the binary is 
\begin{equation}
    l_0 = \frac{\langle \dot J_{\rm{tot}} \rangle}{\langle \dot M \rangle} \simeq 0.709 a_{\rm{B}}^2 \Omega_{\rm{B}} 
\label{eq:torque-constvisc}
\end{equation}
which is similar to the result $l_0 = 0.729a_{\rm{B}}^2 \Omega_{\rm{B}}$ obtained in \haiyang{\citet{2021ApJ...921...71D,2022MNRAS.513.6158D} in which a constant viscosity profile and the torque-free sink prescription are adopted}. 

We employed a second set of diagnostics, measuring different components of angular momentum current due to advection, viscosity and gravitational interaction as shown in Figure \ref{fig:amcurrent-1}. The overlaid yellow horizontal dashed line representing the directly calculated specific torque onto the binary, is in close agreement with the specific angular momentum current represented by the orange line. Hence, we can see that the quasi-steady state of angular momentum transfer both between the disc and the binary and within the CBD is indeed reached.

\textit{Impact of dynamical cooling:} \dong{Figure \ref{fig:torque-subset1} shows the} dependence of these specific torque components and the total specific torque on cooling parameter $\beta$. As $\beta$ becomes larger, the specific angular momentum transfer from the disc to the binary decreases, with a small-magnitude increase around $\beta=2.0$. \im{We also run the case with $\beta=6.0$ (not included) and the decreasing trend remains.} From the fourth panel of Figure \ref{fig:torque-subset1}, it is clear that the binary would experience inspiral \dong{when the equation of state deviates} from locally isothermal. Tracing the origin of this dependence, we can separately analyze the contribution from each torque component: (i) The specific accretion torque is insensitive to the thermodynamics of the CBD. \haiyang{If the binary accretes isotropically, the specific angular momentum of the accreted gas should be the same as that of the binary: $l_{\rm{B}}=q_{\rm{B}} (1+q_{\rm{B}})^{-2} \ a_{\rm{B}}^2\Omega_{\rm{B}}=0.25a_{\rm{B}}^2\Omega_{\rm{B}}$ (see Section 4.2 in \citealt{2020ApJ...901...25D}).} These small differences between the specific accretion torque and contributions from isotropic accretion originate from binary's anisotropic accretion (see Section 2.2 in \citealt{2019ApJ...871...84M} for a detailed explanation). (ii) The specific gravitational torque is the main cause of binary orbital inspiral. Comparing the results from the second and the fourth panels in Figure \ref{fig:torque-subset1}, we see that the gravitational torque outside $r=a_{\rm{B}}$ is responsible for rapid change in total torques. This also suggests that in this simulation setup, the contribution from the mini-discs is insensitive to the cooling timescale. \im{
We speculate that the reason for this is closely related to the morphology change and postpone discussion to Section \ref{subsubsec:constvisc-morphology}.}


\textit{Interpretation on binary orbital evolution:}
Figure \ref{fig:amcurrent-1} and \ref{fig:amcurrent-2} \dong{show the} radial profiles of angular momentum current \dong{for} $\beta \simeq 0$ (locally isothermal) and $0.5$. The former and the latter separately correspond to an outspiraling and an inspiraling binary. The differences between the two are clear: when $\beta=0.5$, \im{the specific gravitational angular momentum current around $r=a_{\rm{B}}$ and the specific viscous angular momentum current at outer radial location are larger}\footnote{$T_{\rm{grav}}$ is larger around $a_{\rm{B}}$, and $\dot J_{\rm{visc}}$ is larger at larger radial location.}, while the specific advection angular momentum current is comparable \haiyang{to} the locally isothermal scenario. The same can be seen in the third panel of Figure \ref{fig:torque-subset1}, where the gravitational torque outside $a_{\rm{B}}$ is smaller when $\beta=0.5$ comparing to $\beta=0.0$. This can be credited to a shallower and smaller cavity when $\beta$ is larger (as illustrated in Figure \ref{fig:morphology-1}, see a further description in Section \ref{subsubsec:constvisc-morphology}).

\begin{figure*}
\begin{center}
\includegraphics[width=0.65\textwidth,trim={0cm 0cm 0cm .0cm},clip]{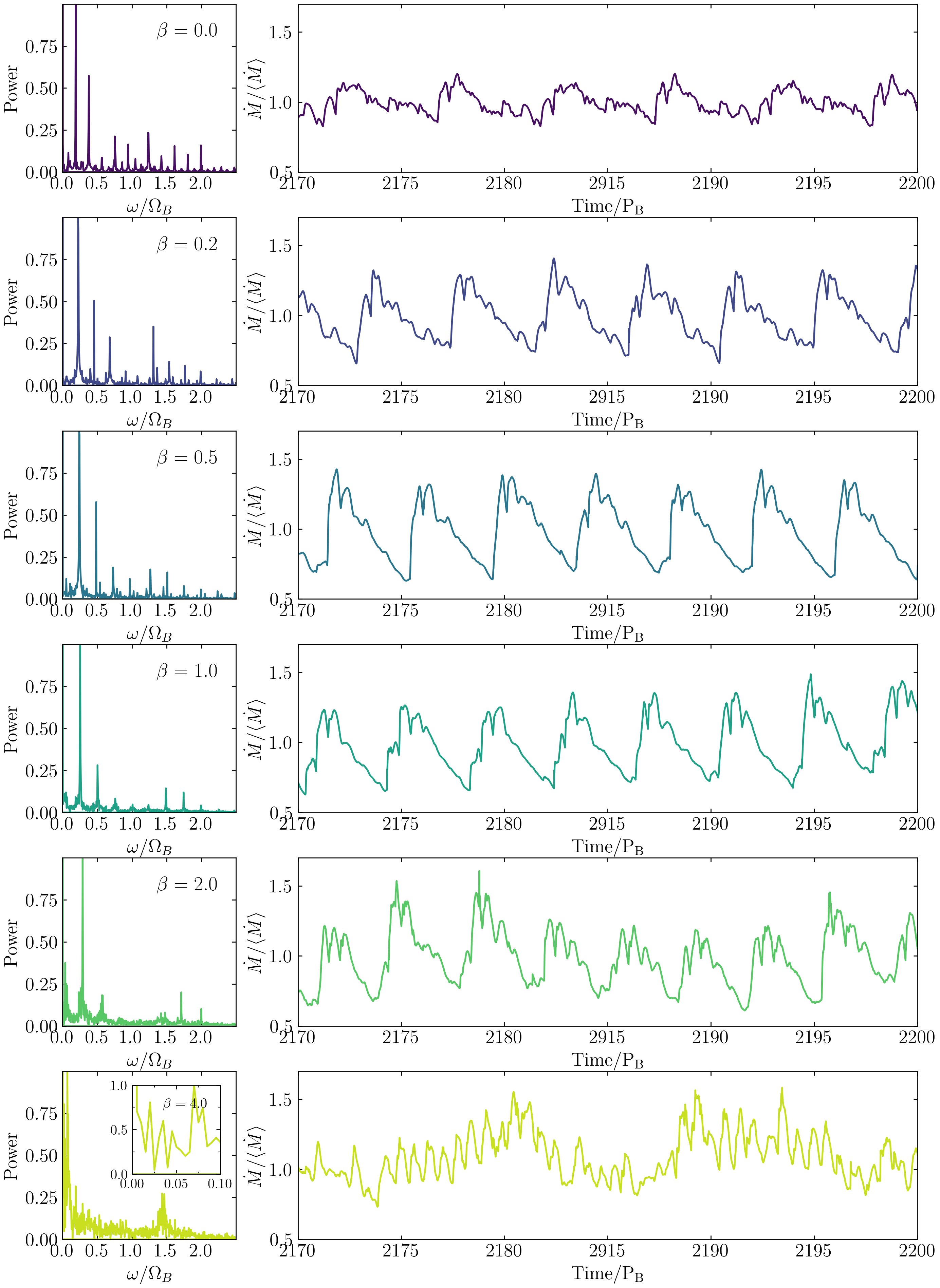}
\end{center}
\vspace{-0.3cm}
\caption{
Accretion variability in the fiducial run subset \texttt{t1n1b-} with constant kinematic viscosity. The first column is the power spectrum of the total accretion rates and the right column is the time series of accretion rate in $2170P_{\rm{B}} - 2200P_{\rm{B}}$. From the top to the bottom, the value of cooling parameter $\beta$ gradually increases following the sequence $[0.0, 0.2, 0.5, 1.0, 2.0, 4.0]$. \im{A zoom-in power spectrum of low-frequency variability when $\beta=4.0$ is embedded in the bottom left panel.}
\label{fig:accvar-1}}
\vspace{-0.3cm}
\end{figure*}


\begin{figure*}
\begin{center}
\includegraphics[width=0.98\textwidth,trim={0cm 0cm 0cm .0cm},clip]{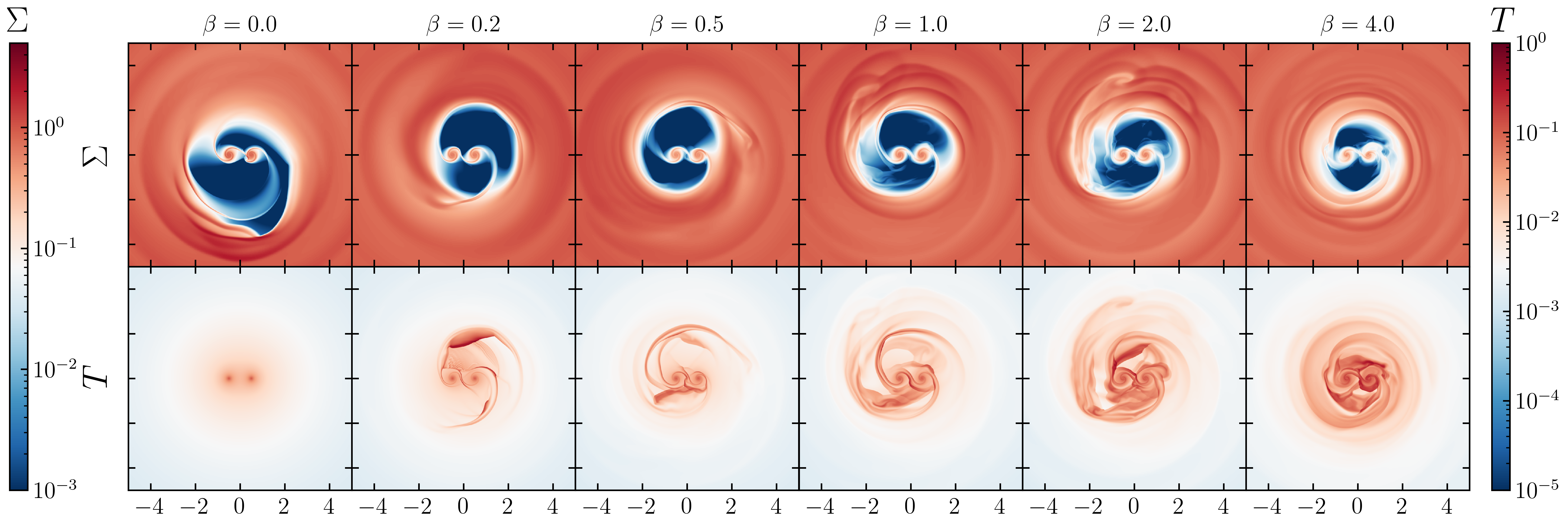}
\end{center}
\vspace{-0.3cm}
\caption{
The density distribution (first row) and the temperature distribution (second row) in the fiducial run subset \texttt{t1n1b-} with constant kinematic viscosity. From the left to the right, the value of $\beta$ increases. 
\label{fig:morphology-1}}
\vspace{-0.3cm}
\end{figure*}

\begin{figure*}
\begin{center}
\includegraphics[width=0.98\textwidth,trim={0cm 0cm 0cm .0cm},clip]{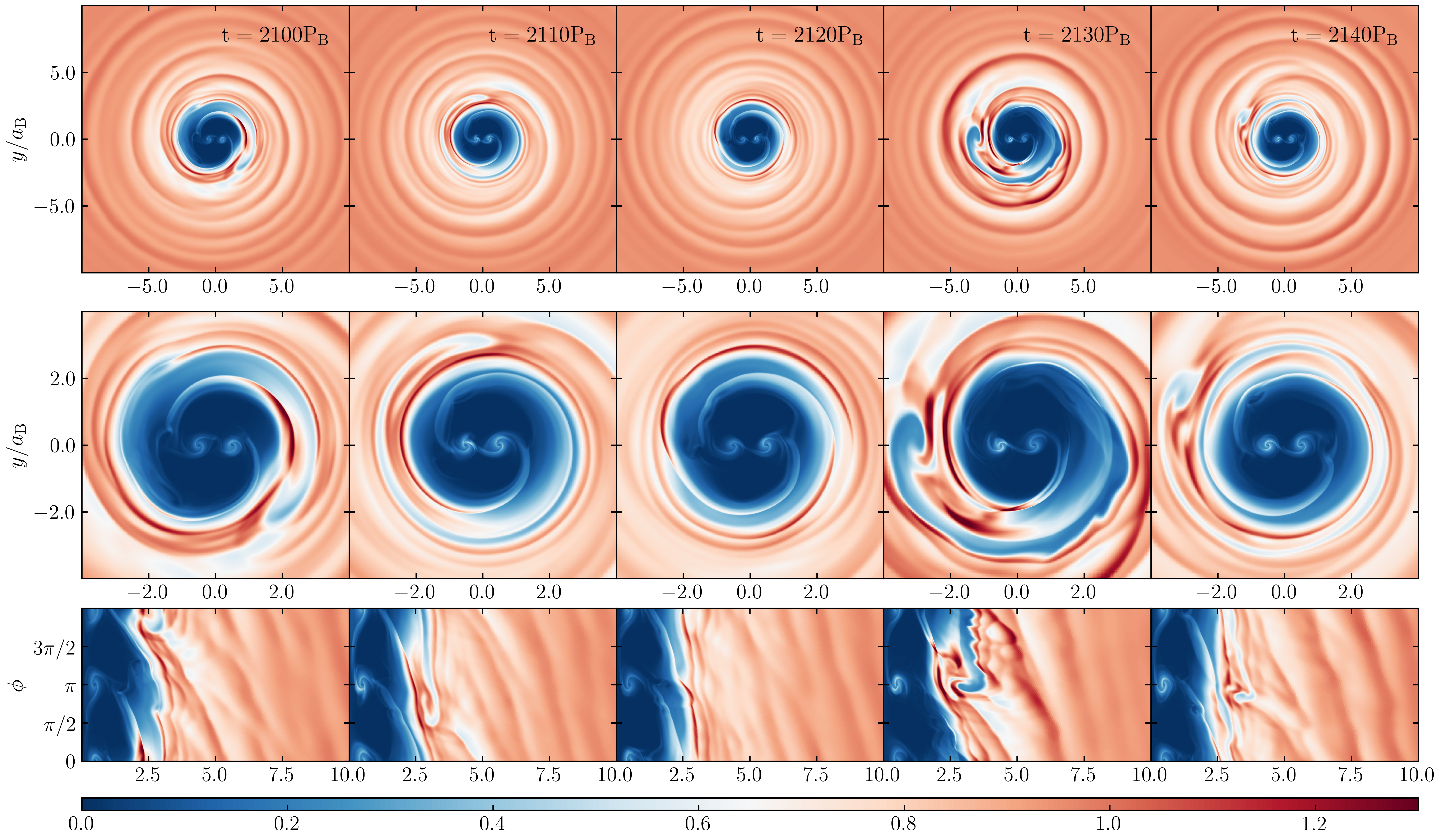}
\end{center}
\vspace{-0.3cm}
\caption{
\im{A series of snapshots of surface density $\Sigma$ with parameterized cooling time $\beta=4.0$ in run \texttt{t1n1b4.0}. The upper two rows are the original simulation snapshots in cartesian geometry $(x, y)$: The panels in the first row include a larger radial range and emphasize the `outer region' of the disk; the panels in the second row are the zoom-in views emphasizing the structure in the `inner region'. In the bottom row shows the interpolated density distribution in $(r, \phi)$, to better illustrate the pitch angles of the spiral density waves.
The 1st and 4th columns has a single major density peak representing an one-arm spiral, and the 3rd column shows the state dominated by two-armed spirals. Other panels show the intermediate state of the transition.
The density is in linear scale, different from the log scale used in other figures, as shown in the colorbar.}
\label{fig:spiral}}
\vspace{-0.3cm}
\end{figure*}

\subsubsection{Accretion Variability}
\label{subsubsec:constvisc-accretion}

To explore the dependence of accretion variability on the cooling time, we keep track of the accretion rate onto each binary component 100 times per orbit. The power spectrum (left column) and the time series (right column) of accretion rate are shown in Figure \ref{fig:accvar-1}, for $\beta \in [0.0, 0.2, 0.5, 1.0, 2.0, 4.0]$ from the top to the bottom. The power spectra are calculated using the time series of accretion rate in the last 200 orbits. They are also normalized by the largest amplitude to show the relative difference.

The magnitude of accretion variability and the characterized frequencies are the two features that we focus on here. (i) The accretion variability is not significantly suppressed with increasing $\beta$ in simulations employing constant viscosity, unlike that of the simulations employing $\alpha-$viscosity in \citet{paper0} and Section \ref{subsubsec:varyvisc-accretion}. The variation of accretion rate is always around $25\%-50\%$ of the average accretion rate. (ii) \dong{Except for the $\beta=4.0$ case, the dominant variability frequency is about $\Omega_{\rm{B}}/5$} This is widely acknowledged as the ``lump-modulated'' accretion variability (e.g., \citealt{2008ApJ...672...83M,2013MNRAS.436.2997D,2017MNRAS.466.1170M}). There are also a series of other periodic variations with higher frequencies which can be seen in the power spectrum. These are also found in previous works (e.g., \citealt{2017MNRAS.466.1170M}). The relatively large peak near the frequency of $\sim 2\Omega_{\rm{B}}$ is related to the accretion streams carrying materials from the CBD to the mini-discs, which periodically repeat themselves every half of the binary orbit. The case of $\beta=4.0$ is worth special care and we postpone the discussion to Section \ref{subsubsec:qpv}.

\haiyang{It is worth noting that, unlike specific torques, accretion variability calculated can be strongly dependent on sink prescriptions, including the type of the sink, sink radius, and sink timescale (as pointed out by \citealt{2014ApJ...783..134F}, \citealt{2017MNRAS.469.4258T}, \citealt{2020ApJ...901...25D} and \citealt{2021ApJ...921...71D}). \im{Therefore,} the validity of our results in this section could be affected if a significantly different sink term is adopted.}
\im{Since our sink depends explicitly on the viscosity, as shown in Eq \ref{eq:tsink}, the dependence of accretion variability on cooling time $\beta$ is qualitatively different when adopting different viscosity choices.}

\begin{figure*}
\begin{center}
\includegraphics[width=0.98\textwidth,trim={0cm 0cm 0cm .0cm},clip]{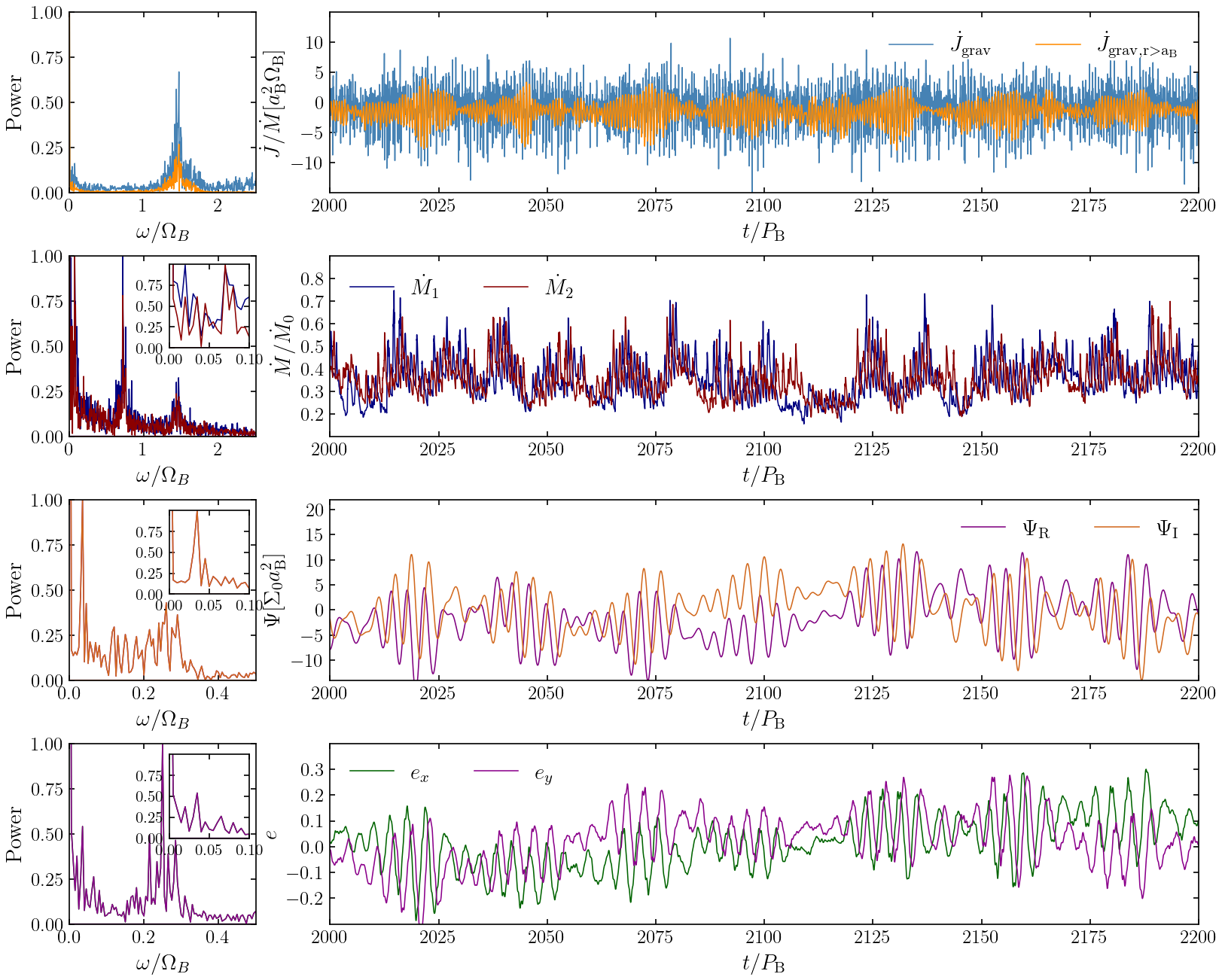}
\end{center}
\vspace{-0.3cm}
\caption{
\im{Power spectrum (left column) and variability (right column) of different diagnostics in fiducial run \texttt{t1n1b4.0} over the period of $[2000, 2200]P_{\rm{B}}$.} From the top to the bottom are (i) the specific angular momentum transfer rate due to gravitational interaction $\dot J_{\rm{grav}}$, (ii) the accretion rate onto each component of the binary $\dot M$, (iii) the mass dipole of the CBD $\Psi$, and (iv) the mass-weighted eccentricity of the cavity $e$. \im{In the first two rows, the individual lines with the same color in the power spectra separately match the cooresponding lines in the time series. In the 3rd and 4th rows, the lines denote the magnitude of mass quadrupole $|\Psi|=(|\Psi_{\rm{R}}|^2+|\Psi_{\rm{I}}|^2)^{1/2}$ and eccentricity $|e|=(|e_{\rm{x}}|^2+|e_{\rm{y}}|^2)^{1/2}$.} The QPV feature is most prominent in the last two panels. The calculated mass dipole and eccentricity have the maximum magnitudes when the single-armed spiral is present and have the minimum magnitudes when the two-armed spirals are present. \im{The power spectra of the mass dipole and eccentricity emphasize the low-frequency component in order to capture the QPV behavior.}
\label{fig:QPV-variability}}
\vspace{-0.3cm}
\end{figure*}

\subsubsection{Disc Morphology}
\label{subsubsec:constvisc-morphology}

Figure \ref{fig:morphology-1} shows a series of snapshots of disc surface density and temperature with increasing cooling timescale $\beta$ from left to right. The binary's gravitational tidal potential clears an eccentric low-density cavity, which is penetrated by narrow accretion streams characterized by azimuthal number $m=1$ or $2$. The high-density lump orbiting around the cavity wall is also prominent. 

As $\beta$ increases, the most significant changes occur in the shape and size of the cavity: the cavity gradually becomes ``shallower'' and more symmetric. These features are also reported in \citet{2022A&A...664A.157S}. ``Shallower'' cavity is also the signal that the gravitational torque truncates the CBD not as effectively as in the locally isothermal case. This results from a hotter cavity with larger $\beta$, which can be seen from the temperature profile in the lower panels. This trend of ``a more symmetric cavity'' is consistent with our findings in \citet{paper0}, despite the fact that in this work the ``twisted spirals'' are no longer present. The absence of ``twisted spirals'' could be originated from the choice of constant viscosity\footnote{As we will show in Section \ref{subsubsec:varyvisc-amtransfer}, the ``twisted spirals'' are present when adopting dynamically varying $\alpha$-viscosity, consistent with our previous finding.}.

\im{From Section \ref{subsubsec:constvisc-amtransfer}, we notice gravitational torques outside $r=a_{\rm{B}}$ is more sensitive to cooling time $\beta$ comparing to those inside $r=a_{\rm{B}}$. The region inside $r=a_{\rm{B}}$ only includes two mini-disks, while the region outside $r=a_{\rm{B}}$ includes most of the low-density cavity, the narrow accretion streams and the circumbinary disk. Thinking in the binary's corotating frame, the two dimensional time-averaged gravitational torque density only depends on the time-averaged disk mass density, since the gravitational interaction is a function of distance only. From Figure \ref{fig:morphology-1}, we notice that as $\beta$ changes, the flow structure only varies largely in the cavity and the circumbinary disks, while the two mini-disks remains largely unaffected. Hence, we speculate that this is the reason for the high sensitivite dependence of the gravitational torque outside $r=a_{\rm{B}}$ on $\beta$.}

\subsubsection{Quasi-Periodic Variation(QPV) in Morphology}
\label{subsubsec:qpv}
We found in Run \texttt{t1n1b4.0} with $\beta=4.0$, the disc morphology exhibits quasi-periodic variability (QPV, for short).
Shown in Figure \ref{fig:spiral}, QPV can be regarded as a periodic transition of disc morphology from single-armed spiral density waves to two-armed spiral density waves. Since this transition happens in different scales simultaneously,  we show the density snapshots at $[2100,2110,2120,2130,2140]P_{\rm{B}}$ of the CBD ($r<10a_{\rm{B}}$) in the first row of Figure \ref{fig:spiral} and the zoom-in figure in the second row. At a certain time, there exists either a large spiral structure spanning from $\sim2a_{\rm{B}}$ to $\sim10a_{\rm{B}}$ and lopsided accretion streams in the disc cavity; or larger-scale two-armed spirals far from the binary and tightly-wounded two-armed accretion streams at the cavity wall. It is worth noticing that the large-scale spiral features are not present in the locally isothermal simulations. The first row of Figure \ref{fig:spiral} shows that the density waves propagate far from the binary, unlike the density waves dissipate severely within $\sim 5a_{\rm{B}}$ in the locally isothermal case (see, e.g., \citealt{2017MNRAS.466.1170M}). 

The large-scale density waves are generated or strengthened by the stream crossing in the vicinity of the binary (similar to \citealt{2019ApJ...875L..21M}). \im{This is one possible reason for the coexistence of different scale spiral features in both the inner and outer region. The quasi-periodic variability arises from the transition between two possible senarios:} (i) As illustrated in the \st{leftmost panels} \im{1st and 4th panel from the left in Figure \ref{fig:spiral}}, the accretion streams connecting to the mini-discs are merged to form a ``lump''-like high-density structure at the cavity wall. And this lump can drive the single-armed spiral formation in the large scale. (ii) When the morphology in the disc cavity is \im{approximately axi}symmetric, as in the 3rd panel, the lump at the cavity wall is replaced by two-armed, tightly wounded spiral density waves. The absence of the lump leads to more symmetric structure in the large scale (with azimuthal number $m=2$) as well. This quasi-periodic variability has a period of $\sim 30 P_{\rm{B}}$ as shown in Figure \ref{fig:accvar-1}, \ref{fig:spiral}, and \ref{fig:QPV-variability}. \im{In particular, the radially integrated disk eccentricity and mass dipole best illustrute the quasi-periodic feature as opposed to the accretion variability(see, e.g., the low-frequency peaks in the zoom-in power spectrum in Figure \ref{fig:QPV-variability}).}

The existence of this quasi-periodic variability has a large impact on accretion variability and angular momentum transfer between the disc and the binary. \dong{The panels in} Figure \ref{fig:QPV-variability} show the rate of specific angular momentum transfer from the CBD to the binary, accretion rate onto each binary component, mass quadrupole of the disc and the mass-weighted eccentricity of the cavity in the period of $[2000, 2200] P_{\rm{B}}$. There exists a clear correlation between these four quantities. Especially, the mass accretion rate exhibits long-term variability, which is not reported in the previous 2D hydrodynamical simulations. 

Perhaps more curiously, \dong{similar features have been reported} in 3D MHD simulations of \citet{2015ApJ...807..131S}\dong{, which use} a globally isothermal equation of state with sound speed $0.1\Omega_{\rm{B}}a_{\rm{B}}$, and with viscosity provided by MRI turbulence. And our simulations that employ dynamically varying $\alpha-$viscosity, which will be discussed in Section \ref{subsec:varyvisc}, show similar features when the disk is hot enough ($\beta = 4.0$). Further work is needed investigate the origin of this phenomenon.

\subsection{Results for Dynamically Varying Viscosity}
\label{subsec:varyvisc}

Similar to Section \ref{subsec:constvisc}, we follow our standard suite of analysis for the simulation subset \texttt{t1n3b-} with dynamically varying $\alpha-$viscosity. In this series of runs, the viscosity profile is dynamically influenced by the temperature profile, which is set by the balance between viscous heating and $\beta-$cooling. Several differences are expected: with higher temperature in the cavity and consequently larger viscosity, the disc self-adjusts faster towards a quasi-steady state. It further influences the angular momentum transfer rate and binary's accretion variability.

\subsubsection{Angular Momentum Transfer and disc Morphology}
\label{subsubsec:varyvisc-amtransfer}

\begin{figure}
\begin{center}
\includegraphics[width=0.45\textwidth,trim={0cm 0cm 0cm .0cm},clip]{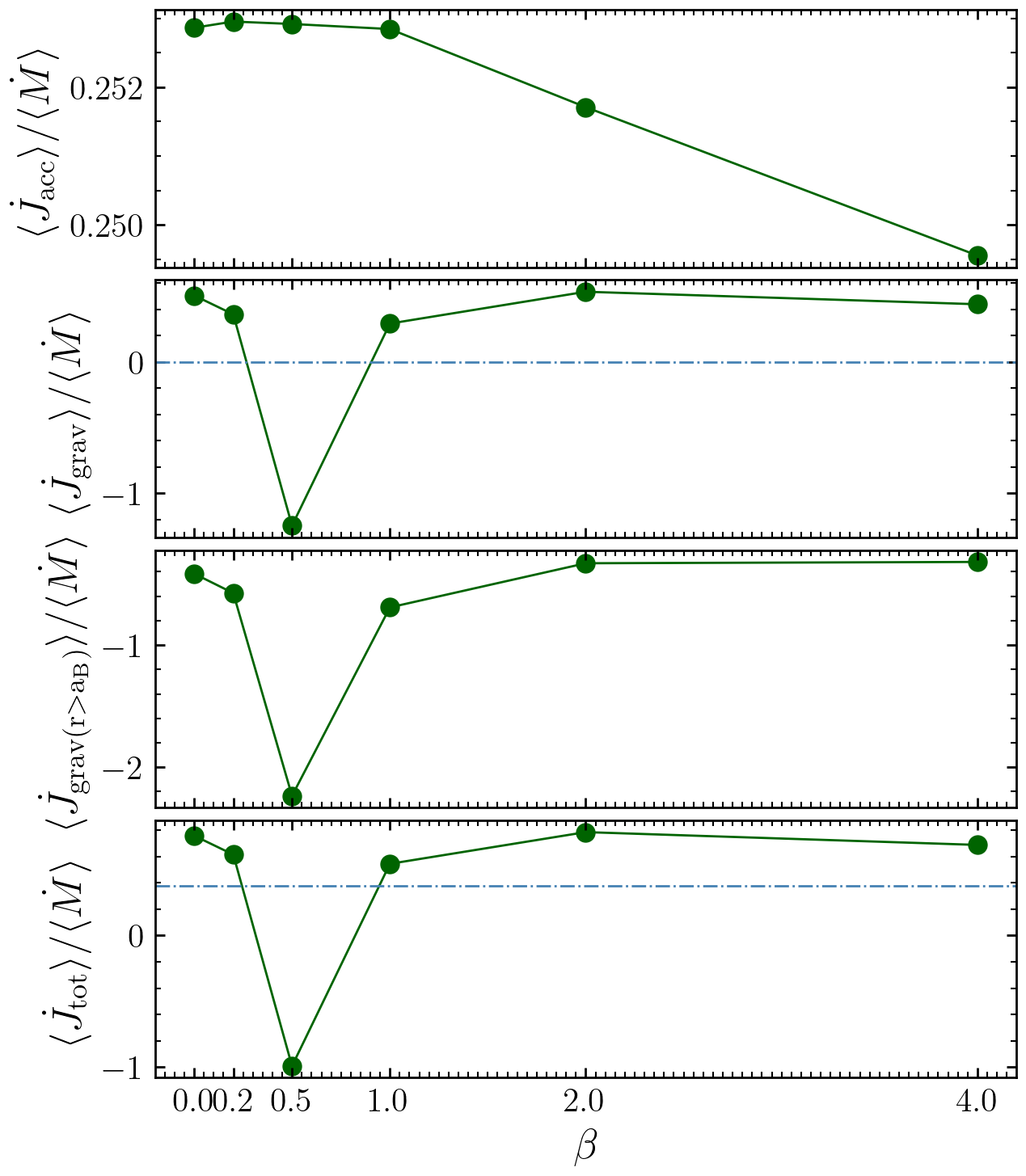}
\end{center}
\vspace{-0.3cm}
\caption{Same as Figure \ref{fig:torque-subset1}, with the four panels showing how different torque components and the total torque vary with $\beta$ in fiducial run subset \texttt{t1n3b-} (with dynamically varying viscosity).
\label{fig:torque-subset2}}
\vspace{-0.3cm}
\end{figure}

\begin{figure*}
\begin{center}
\includegraphics[width=0.98\textwidth,trim={0cm 0cm 0cm .0cm},clip]{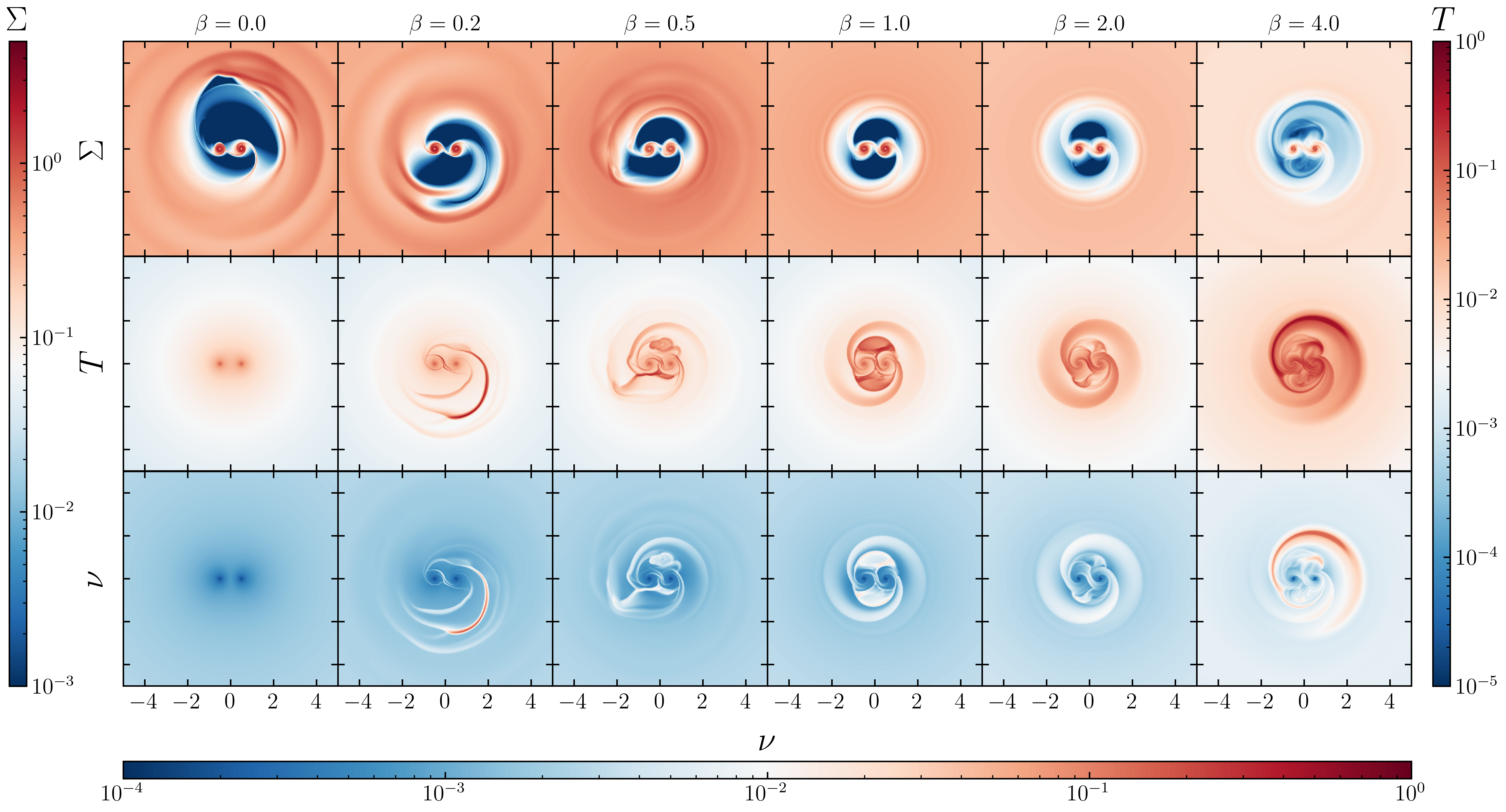}
\end{center}
\vspace{-0.3cm}
\caption{
Similar to Figure \ref{fig:morphology-1}, showing the density distribution (first row), the temperature distribution (second row) and the viscosity distribution (third row) in the fiducial run subset \texttt{t1n3b-} with dynamically varying $\alpha-$viscosity.
\label{fig:morphology-2}
}
\vspace{-0.3cm}
\end{figure*}

\begin{figure}
\begin{center}
\includegraphics[width=0.45\textwidth,trim={0cm 0cm 0cm .0cm},clip]{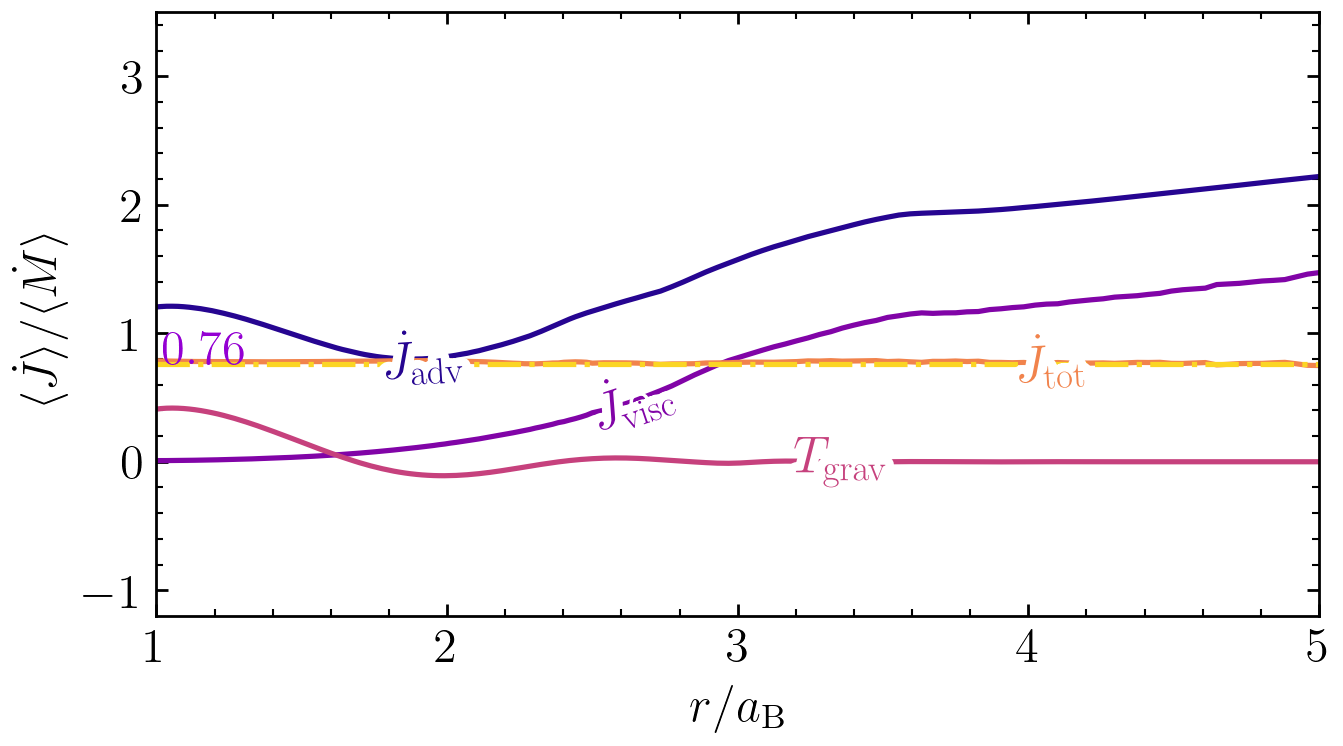}
\end{center}
\vspace{-0.3cm}
\caption{
Similar to Figure \ref{fig:amcurrent-1}, \dong{but using the dynamically varying $\alpha-$viscosity}. The corresponding simulation ID is \texttt{t1n3b0.0}.
\label{fig:amcurrent-3}
}
\vspace{-0.3cm}
\end{figure}

\begin{figure}
\begin{center}
\includegraphics[width=0.45\textwidth,trim={0cm 0cm 0cm .0cm},clip]{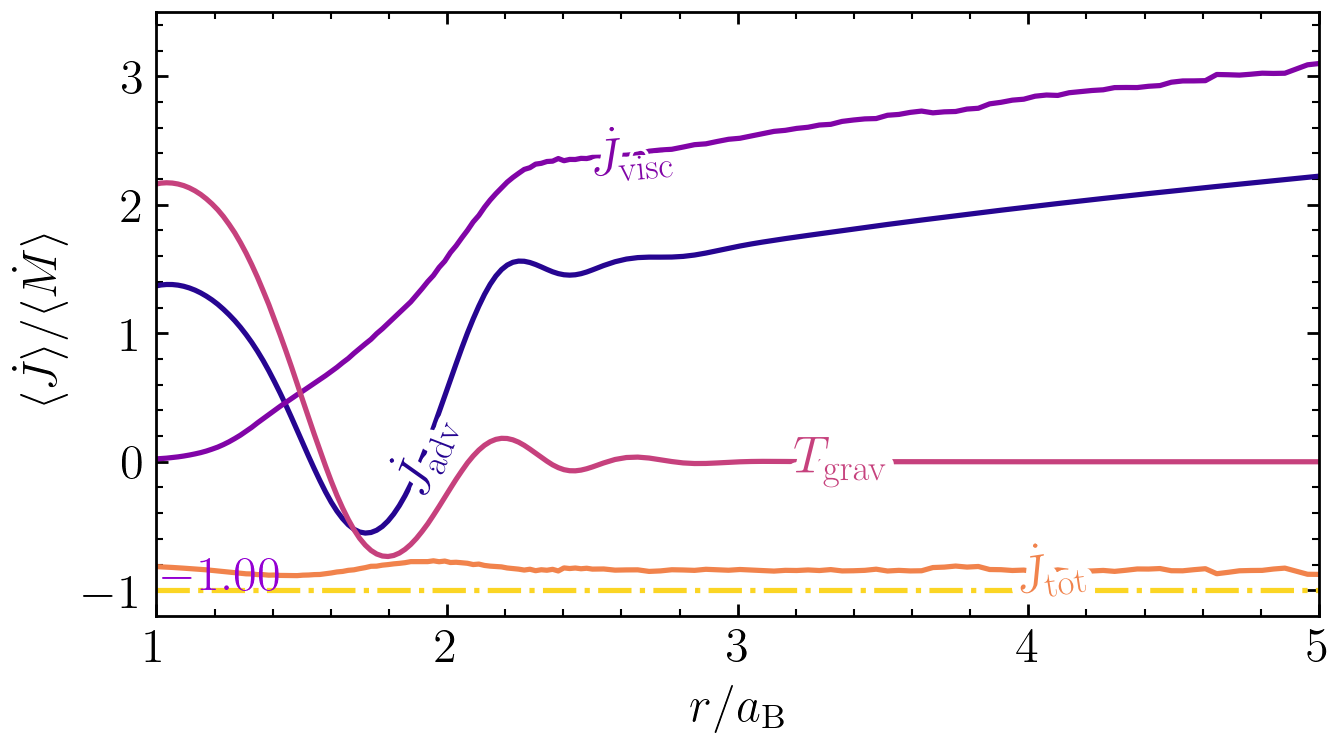}
\end{center}
\vspace{-0.3cm}
\caption{
Similar to Figure \ref{fig:amcurrent-3}, but with cooling parameter $\beta=0.5$. In comparison to Figure \ref{fig:amcurrent-3}, the gravitational torque and viscous torque are significantly larger, resulting in a significantly smaller total angular momentum current. The corresponding simulation ID is \texttt{t1n3b0.5}.
\label{fig:amcurrent-4}}
\vspace{-0.3cm}
\end{figure}

\begin{figure}
\begin{center}
\includegraphics[width=0.45\textwidth,trim={0cm 0cm 0cm .0cm},clip]{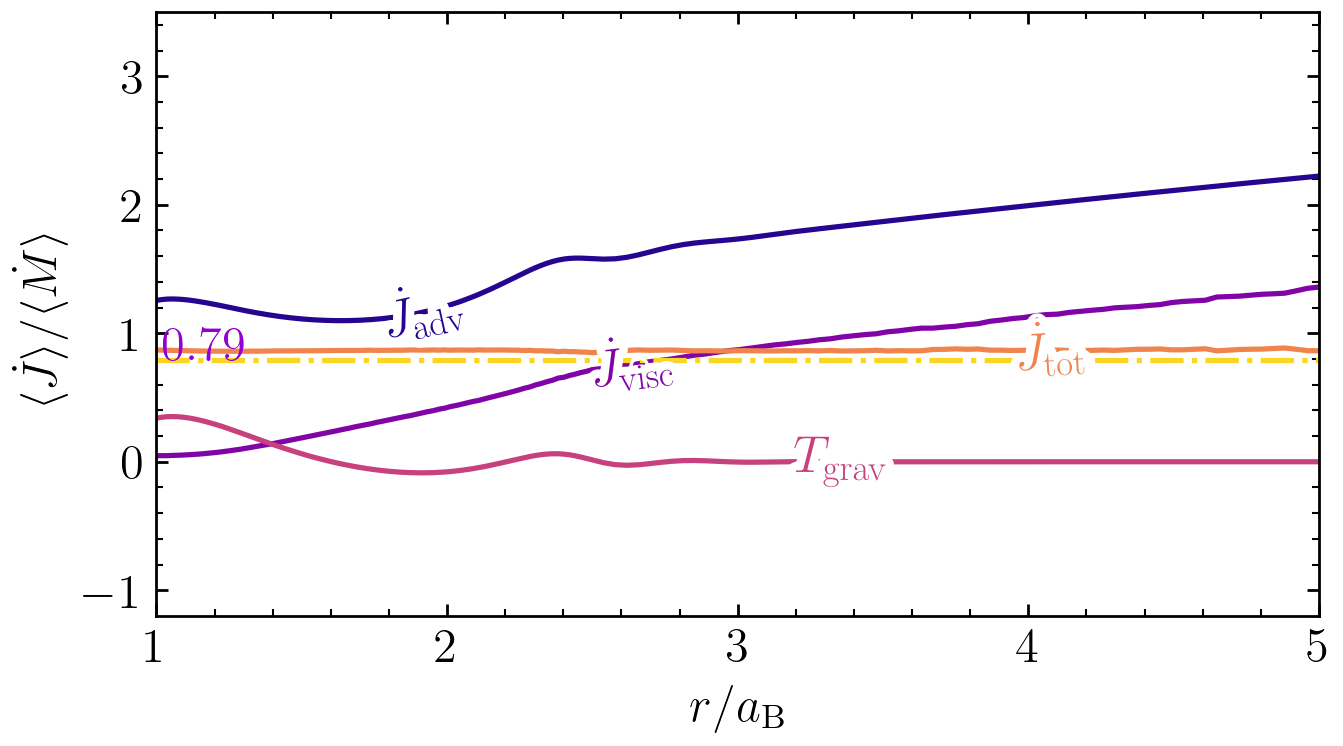}
\end{center}
\vspace{-0.3cm}
\caption{
Similar to Figure \ref{fig:amcurrent-3}, but with cooling parameter $\beta=2.0$. Different torque components resemble those in the locally isothermal case. The corresponding simulation ID is \texttt{t1n3b2.0}.
\label{fig:amcurrent-5}}
\vspace{-0.3cm}
\end{figure}

Again, we summarize the dependence of \dong{various} specific torque components on the cooling parameter $\beta$ in Figure \ref{fig:torque-subset2}. The total angular momentum current per unit mass through the CBD in the locally isothermal limit is illustrated in Figure \ref{fig:amcurrent-3}, in which case the specific total torque is given by
\begin{equation}
	\frac{\langle \dot J _{\rm{tot}} \rangle}{\langle \dot M \rangle} \simeq 0.759 a_{\rm{B}}^2\Omega_{\rm{B}}
\label{eq:torque-varyvisc}
\end{equation}
This value is close to the result for the constant-viscosity case (Equation \ref{eq:torque-constvisc}), and similar to \citet{2022MNRAS.513.6158D} ($l_0 = 0.767a_{\rm{B}}^2 \Omega_{\rm{B}}$) and is somewhat larger than the result ($l_0 = 0.723a_{\rm{B}}^2 \Omega_{\rm{B}}$) of \citet{2019ApJ...875...66M}. 
\im{The differences of these two results is mainly from the choice of sink prescription: the former uses the torque-free sink while the latter uses the standard sink.}

\textit{Impact of dynamical cooling:} From Figure \ref{fig:torque-subset2}, we \dong{see that} the angular momentum transfer between the disc and the binary \dong{depends on $\beta$ in a non-monotonic way, with the minimum of $\langle \dot J _{\rm{tot}} \rangle/\langle \dot M \rangle$}
 achieved at $\beta=0.5$. This trend is similar to our previous findings in \citet{paper0}. Similar to Section \ref{subsec:constvisc}, it is useful to decompose the total specific torque to accretion torque and gravitational torque. We find: (i) The specific accretion torque is insensitive to the thermodynamics, with a smaller variation compared to the case with constant viscosity; the specific accretion torque is very close to the specific torque of the binary $l_{\rm{B}}=0.25a_{\rm{B}}^2\Omega_{\rm{B}}$. (ii) The specific gravitational torque, and more specifically, the gravitational torque outside of $r=a_{\rm{B}}$, is responsible to the ``v'' shape transition as $\beta$ increases. 

\textit{Interpretation on binary orbital evolution:} From the morphology of the runs shown in Figure \ref{fig:morphology-2}, it is easy to notice a prominent transition in the CBD cavity from eccentric to circular when $\beta$ changes from $0.2$ to $1.0$. When $\beta = 0.5$ (the third column in Figure \ref{fig:morphology-2}), the disk materials at the edge of the cavity wall exhibit ``twisted'' features and leave a trailing spiral extending to $r\approx 5a_{\rm{B}}$. \im{The accretion variability is solely provided by the variability of the ``twisted accretion stream''.} \dong{As $\beta$ increases further,  the morphology of both the CBD and the accretion streams} roughly stay fixed and their patterns corotate with the binary: the cavity is penetrated by two-armed, tidally induced spirals directly connecting the binary and the CBD. The reason for the ``v'' shaped relation in the angular momentum transfer rate is qualitatively similar to our explanation in \citet{paper0}: From Figure \ref{fig:torque-subset2}, we see that the major variation in total gravitational torques is originated from the region $r>a_{\rm{B}}$, which means that it is the contribution from the accretion streams in the cavity and the outer CBD that sensitively depends on the cooling time. Meanwhile, we show the radial dependence of different angular momentum current components of run \texttt{t1n3b0.5} \dong{(corresponding to the case of binary inspiral in Figure \ref{fig:torque-subset2})} in Figure \ref{fig:amcurrent-4}. \dong{We see that} the angular momentum current from gravitational interaction increases significantly in the radial range of $[1.0,1.7]a_{\rm{B}}$, which corresponds to the location of accretion streams connecting the binary and the CBD. The specific angular momentum current due to viscous transport is also larger than that due to advection. \im{For comparison purpose, we also show the torque components of run \texttt{t1n3b2.0} in Figure \ref{fig:amcurrent-5}. The overall radial profile of various torques resemble those in the locally isothermal case, despite the cavity is mostly circular in $\beta=2.0$ case.} Therefore, the ``v'' shaped feature \dong{in Figure \ref{fig:torque-subset2} likely stems from} the cavity's morphology transitioning from eccentric to circular. 
\haiyang{It is worth noticing that the specific angular momentum transfer rate onto the binary (yellow dash-dotted line) slightly deviates from that across the CBD (orange solid line) in run \texttt{t1n3b0.5} as shown in Figure \ref{fig:amcurrent-4}. The difference is most likely resulted from the contribution from pressure and the finite size of the sink cell (see, e.g., \citealt{2022MNRAS.517.1602L}). }

\subsubsection{Accretion Variability}
\label{subsubsec:varyvisc-accretion}

\begin{figure*}
\begin{center}
\includegraphics[width=0.65\textwidth,trim={0cm 0cm 0cm .0cm},clip]{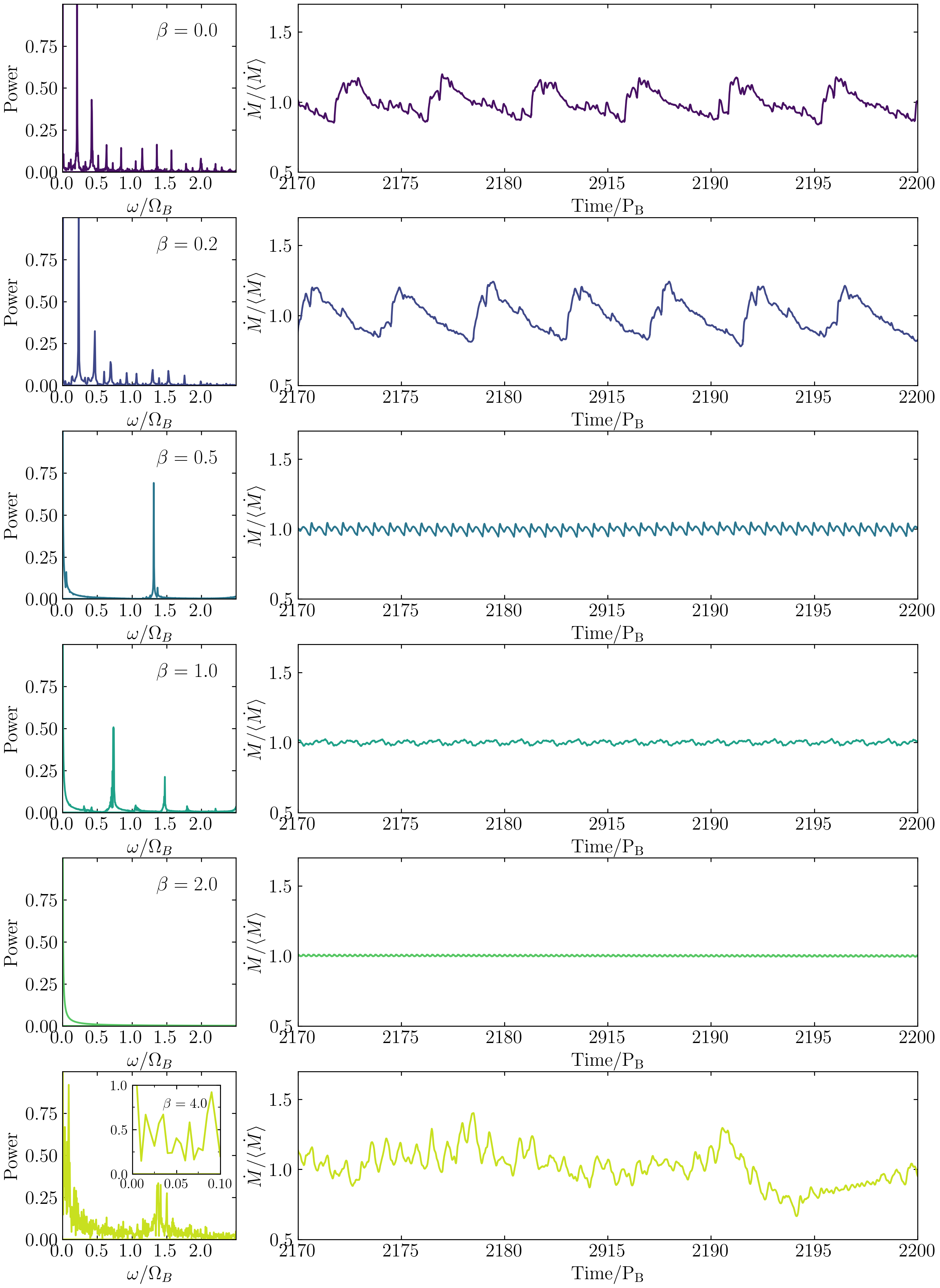}
\end{center}
\vspace{-0.3cm}
\caption{
Similar to Figure \ref{fig:accvar-1}, showing the accretion variability in the fiducial run subset \texttt{t1n3b-} with dynamically varying $\alpha-$viscosity. \im{And a zoom-in power spectrum of low-frequency variability when $\beta=4.0$ is embedded in the bottom left panel.}
\label{fig:accvar-2}}
\vspace{-0.3cm}
\end{figure*}

Similar to the results reported in Section \ref{subsubsec:constvisc-accretion}, we analyze the accretion behaviour of the binary. The normalized power spectrum (left column) and the time series (right column) are shown in Figure \ref{fig:accvar-2}. 

We first focus on the magnitude of accretion variability. It is clear that as $\beta$ increases from $0.0$ to $2.0$ (from top to bottom), the magnitude of accretion rate variability is gradually suppressed. This result is consistent with our findings in \citet{paper0}. The suppression in accretion variability could be traced back to the suppression of variability in the disc morphology shown in Figure \ref{fig:morphology-2}. 

The characterized frequencies of accretion \dong{variability has a complex dependence on $\beta$}. When $\beta$ is small (i.e., $\beta=[0.0, 0.2]$), the accretion behaviour is similar to the locally isothermal case: There are two  characteristic frequencies: the ``lump''-induced, relatively long-term accretion variability \dong{at about} $\Omega_{\rm{B}}/5$, and the variability related to the binary orbital motion and motion of the accretion streams at $2\Omega_{\rm{B}}$. When $\beta=0.5$, the only dominant frequency is $\sim 1.3\Omega_{\rm{B}}$. This is also similar to our previous results \citep{paper0}. The high-density lump disappears and the dominant frequency of accretion is associated with the spiral accretion streams. When $\beta \gtrsim 1.0$, the accretion process is much more stable, corresponding to the transition in disc morphology at $\beta=1.0$ shown in Figure \ref{fig:morphology-2}. Materials are directly transferred from CBD to the binary through the two-armed spirals. We also note that the accretion variability could depend on the sink prescription.This is because the CBD is in phase corotating with the binary and the variabilities associated with the disc morphology changes are efficiently suppressed.  

Similar to \dong{the constant-viscosity} simulations, it is worth noticing that the magnitude of the accretion variability \dong{is larger for} $\beta=4.0$ \dong{than for} $\beta=2.0$. The prominent frequency that characterizes this variability is similar to that found in Section \ref{subsubsec:qpv}. This shows that quasi-periodic variability also exists in simulations \texttt{t1n3b-} adopting dynamically varying $\alpha-$viscosity.

\subsection{Discussion: Parameter Survey}
\label{subsection:survey}

\dong{Our results presented} in section \ref{subsec:constvisc} and \ref{subsec:varyvisc} are based on the two subsets of the simulations in the fiducial runs. \dong{Here} we discuss additional results included in the parameter survey in Table \ref{tab:simulation-parameters-maintext}. There are 4 subsets of simulations: (i) \texttt{t2n2b-}, with axisymmetric \dong{(equilibrium)} temperature profile and axisymmetric, dynamically varying $\alpha-$viscosity; (ii) \texttt{t2n1b-}, with axisymmetric temperature profile and constant viscosity; (iii) \texttt{t1n3b-}*, with the temperature profile centered on each component of the binary and fixed $\alpha-$viscosity; and (iv) \texttt{t2n2b-}*, with axisymmetric temperature profile and axisymmetric fixed $\alpha-$viscosity. \dong{We can} categorize the 4 different subsets along with 2 in the fiducial runs into two types: (i) simulations having a fixed viscosity profile or (ii) a dynamically varying viscosity profile. The former includes \texttt{t1n1b-}, \texttt{t2n1b-}, \texttt{t1n3b-}*, and \texttt{t2n2b-}*, while the latter includes \texttt{t1n3b-} and \texttt{t2n2b-}. As we will show below, similar dependences between the properties, including CBD morphology and angular momentum transfer rate, and cooling time $\beta$ are found within each category. 

We show the inner CBD morphology, temperature distribution (and viscosity distribution if adopting dynamically varying $\alpha-$viscosity) in the 2nd (3rd) rows of each subset in Figure \ref{fig:morphology-6}, \ref{fig:morphology-4}, \ref{fig:morphology-5} and \ref{fig:morphology-7} corresponding to run \texttt{t2n1b-}, \texttt{t2n2b-}, \texttt{t1n3b-}* and \texttt{t2n2b-}*. For simulations adopting a fixed viscosity profile, as the cooling time $\beta$ increases, the CBD cavity gradually becomes \haiyang{shallower} and more circular. Meanwhile, the high-density lump at the cavity wall gradually becomes a series of spirals. The temperature inside the CBD cavity also increases with increasing $\beta$. The morphology of temperature distribution becomes more  closely coupled to the surface density profile. When $\beta=4.0$, a quasi-periodic variability as described in Section \ref{subsubsec:qpv} always appears. For simulations adopting a dynamically varying $\alpha-$viscosity, the CBD cavity quickly circularize as $\beta$ increases and the CBD has the same pattern speed as the rotating speed of the binary. When the inner disc is hot enough, the corotating disc pattern breaks up and the disc again becomes asymmetric/eccentric, at the same time exhibiting quasi-periodic variability. \haiyang{This can be originated from the fact that eccentric modes in the CBD are efficiently damped when the cooling timescale is close to local Keplerian timescale. We will analyse this effect analytically in \citet{paper2}. }



\im{The relationship between the angular momentum transfer rate and the cooling time parameter $\beta$ can vary significantly depending on whether a fixed viscosity profile is used or not.
When a fixed viscosity profile is employed, the angular momentum transfer rate gradually decreases as $\beta$ increases, as illustrated in Figure \ref{fig:torque-set1}.
However, when a dynamically varying $\alpha$-viscosity is adopted, the relationship between the angular momentum transfer rate and $\beta$ exhibits a non-monotonic, "v"-shaped pattern, as demonstrated in Figure \ref{fig:torque-set2}.
These observed trends generally align with the findings discussed in Section \ref{subsubsec:constvisc-amtransfer} and \ref{subsubsec:varyvisc-amtransfer}. It is also worth noting that employing a constant $\alpha$-viscosity is not in line with physical expectations. Hence, runs including \texttt{t1n3b-} and \texttt{t2n2b-} are presented solely for the purpose of comparison.}

By showing the dependence of various diagnostics on cooling time, we find that although different choices of temperature and viscosity profiles can affect the details of the results, the overall trend of this dependence can be simply categorized into two kinds. This parameter survey shows that our results introduced in Section \ref{subsec:constvisc} and \ref{subsec:varyvisc} are not artifacts of specific temperature and viscosity choices: it is a general result that whether or not explicitly including local temperature in viscosity profile will determine the dependence of the disc properties and the angular momentum transport on the cooling time.

\section{Summary}
\label{section:summary}

In this work, we have conducted a suite of 2D hydrodynamical simulations of circumbinary \dong{accretion} in cartesian geometry, focusing on the impacts of viscous heating and dynamical cooling of the gas. We limit our study to equal mass binaries on fixed circular orbits at the center of CBDs with \dong{the (equilibrium)} disc aspect ratio $h=0.1$. To separate the effect of viscous heating and dynamical cooling, we conducted simulations with static, constant kinematic viscosity, and \dong{dynamically varying} $\alpha-$viscosity. The simulation results are discussed concentrating on \dong{the} angular momentum transfer between the disc and the binary, the accretion variability and the disc morphology. This work is the extension of our recent paper \citep{paper0}, which studied the same problem in a cylindrical coordinate and excised the central simulation domain $(r<a_{\rm{B}})$.

Same as \citet{paper0}, our simulations use idealized treatment of thermodynamics by assuming cooling time $t_{\rm{cool}}$ to be a constant fraction of the local dynamical timescale. All of our simulations have reached a quasi-steady state, \dong{such that the time-averaged mass flux and angular momentum flux across the CBD equal the mass and angular momentum transfer rates to the binary.} The key results of our simulations \dong{are} summarized \dong{below}. Perhaps most importantly, we found that when the CBD's equation of state deviates from the locally isothermal limit, all results can sensitively depend on the viscosity prescription, especially whether the viscosity profile is fixed or dynamically varying with local temperature.

(i) Angular momentum transfer: As the parameterized cooling timescale $\beta$ increases, the specific angular momentum transferred from the disc to the binary generally decreases when adopting constant viscosity \dong{(see Figure \ref{fig:torque-subset1})}; even a slight deviation from the locally isothermal limit (i.e., $\beta=0.2$) could result in orbital inspiral. When adopting \dong{the dynamically varying} $\alpha-$viscosity, the specific angular momentum received by the binary \dong{depends non-monotonically on} $\beta$. Similar to our previous work \citep{paper0}, for most $\beta$-values, the binary will experience orbital expansion; the binary could shrink in a small window \dong{around $\beta \simeq 0.5$}.

(ii) Accretion variability: As the cooling time becomes longer, the magnitude and characterized frequencies of accretion variability are barely affected when employing constant viscosity \dong{(see Figure \ref{fig:accvar-1})}. When employing $\alpha-$viscosity, the accretion variability will be gradually suppressed \dong{as $\beta$ increases (see Figure \ref{fig:accvar-2})}. On the other hand, when $\beta=4.0$, the magnitude of the accretion variability is similar to the locally isothermal limit, and exhibits quasi-periodic features with a period of $\sim 30P_{\rm{B}}$ \dong{(see Figure \ref{fig:QPV-variability})}.

\im{\textbf{Rewritten:} (iii) Disc Morphology: With a longer cooling time, the morphology of the inner disc will generally become more symmetric.  When adopting $\alpha$-viscosity, the disc will show a transition: the morphology is similar to the locally isothermal case when $\beta \lesssim 1.0$, and show a fixed pattern corotating with the binary when $\beta \gtrsim 1.0$. And when the inner disc is hot enough (i.e., $\beta=4.0$), the disc will exhibit quasi-periodic variability (QPV) with a period of $\sim 25P_{\rm{B}}$. }


Comparing with our recent work \citep{paper0}, we \dong{have} made a large improvement in this work by resolving the region within \dong{the binary cavity}. There are \dong{several obvious caveats of work}. Firstly, we focus on CBDs with ``equilibrium'' disc aspect ratio $h=0.1$ and equal-mass, circular binaries. The orbital evolution of the binary can sensitively depend on other binary or disc parameters. Secondly, our simulations are carried out in 2D. Only 3D hydrodynamical simulations can capture the possible deflection of density waves perpendicular to the disc midplane. More importantly, 3D MHD simulations which capture MRI turbulence can provide information regarding binary orbital evolution more realistically. Finally, our treatment of the thermodynamics of the \dong{gas} is still over-simplified. \im{The physical cooling mechanism will need a location-dependent cooling parameter $\beta$ in the cavity and the CBD.} We will aim to include radiative cooling and potentially radiation transport \dong{to attain a more realistic picture of CBD accretion}.

\section*{Acknowledgements}

We thank the anonymous referee for the insightful comments and suggestions. We also thank Gordon Ogilvie for helpful discussion, as well as the KITP program ``Bridging the Gap: Accretion and Orbital Evolution in Stellar and Black Hole Binaries'' for helpful communications. This work is supported by the Dushi fund at Tsinghua University, and in part by the National Science Foundation under Grant No. NSF PHY-1748958. Numerical simulations are conducted in the Orion cluster at Department of Astronomy, Tsinghua University, and in TianHe-3 (A) at National Supercomputer Center in Tianjin, China.

\section*{Data Availability}

 \im{The data obtained in our simulations can be made available on
reasonable request to the corresponding author.}



\bibliographystyle{mnras}
\bibliography{CBD-simulation} 



\appendix

\section{Parameter Survey: Temperature Profile and Viscosity Prescription}
\label{appsec:parameter}

\begin{figure*}
\begin{center}
\includegraphics[width=0.98\textwidth,trim={0cm 0cm 0cm .0cm},clip]{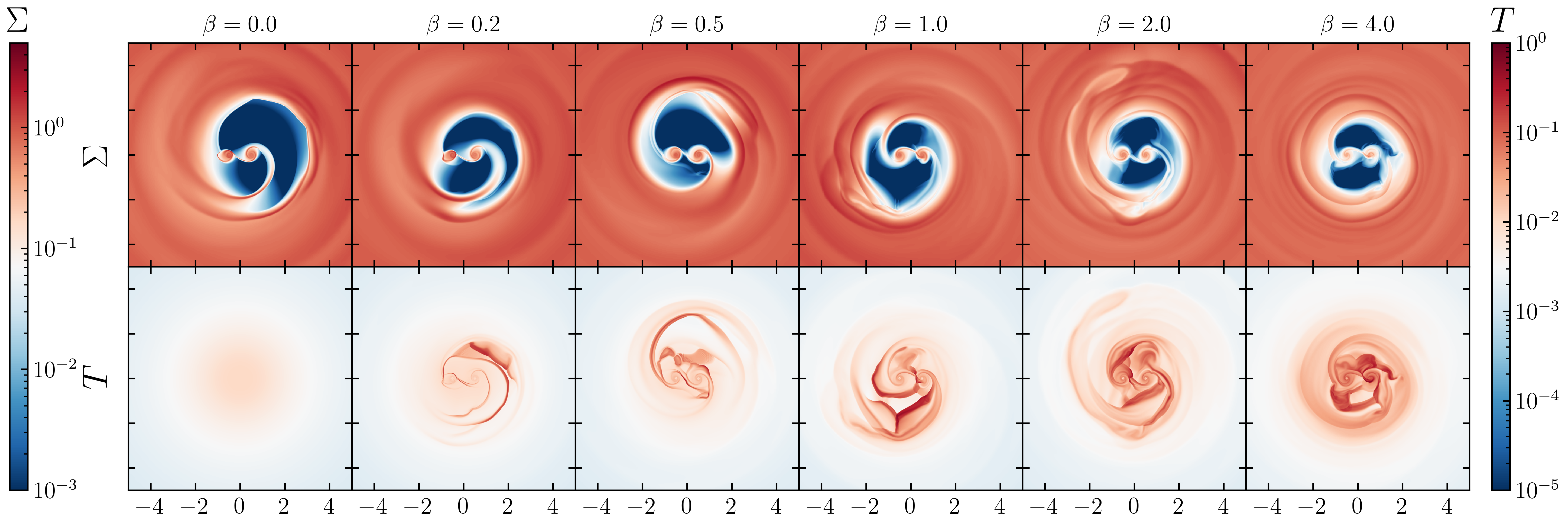}
\end{center}
\vspace{-0.3cm}
\caption{
Similar to Figure \ref{fig:morphology-1}, the density distribution (first row) and the temperature distribution (second row) in parameter survey subset \texttt{t2n1b-} with constant kinematic viscosity.
\label{fig:morphology-6}
}
\vspace{-0.3cm}
\end{figure*}

\begin{figure*}
\begin{center}
\includegraphics[width=0.98\textwidth,trim={0cm 0cm 0cm .0cm},clip]{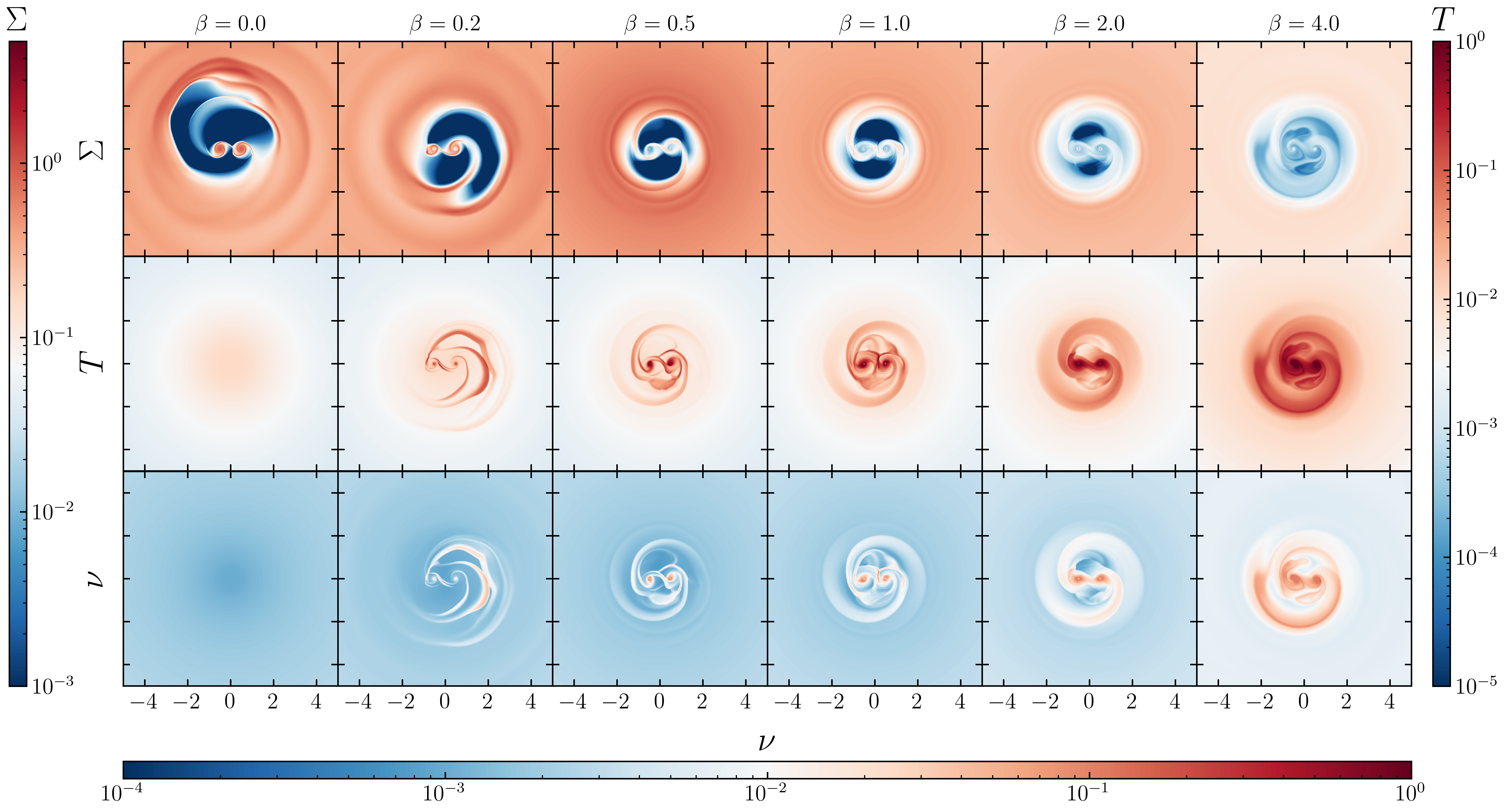}
\end{center}
\vspace{-0.3cm}
\caption{
Similar to Figure \ref{fig:morphology-2}, the density distribution (first row), the temperature distribution (second row) and the viscosity distribution (third row) in parameter survey subset \texttt{t2n2b-} with dynamically varying $\alpha-$viscosity.
\label{fig:morphology-4}
}
\vspace{-0.3cm}
\end{figure*}

\begin{figure*}
\begin{center}
\includegraphics[width=0.98\textwidth,trim={0cm 0cm 0cm .0cm},clip]{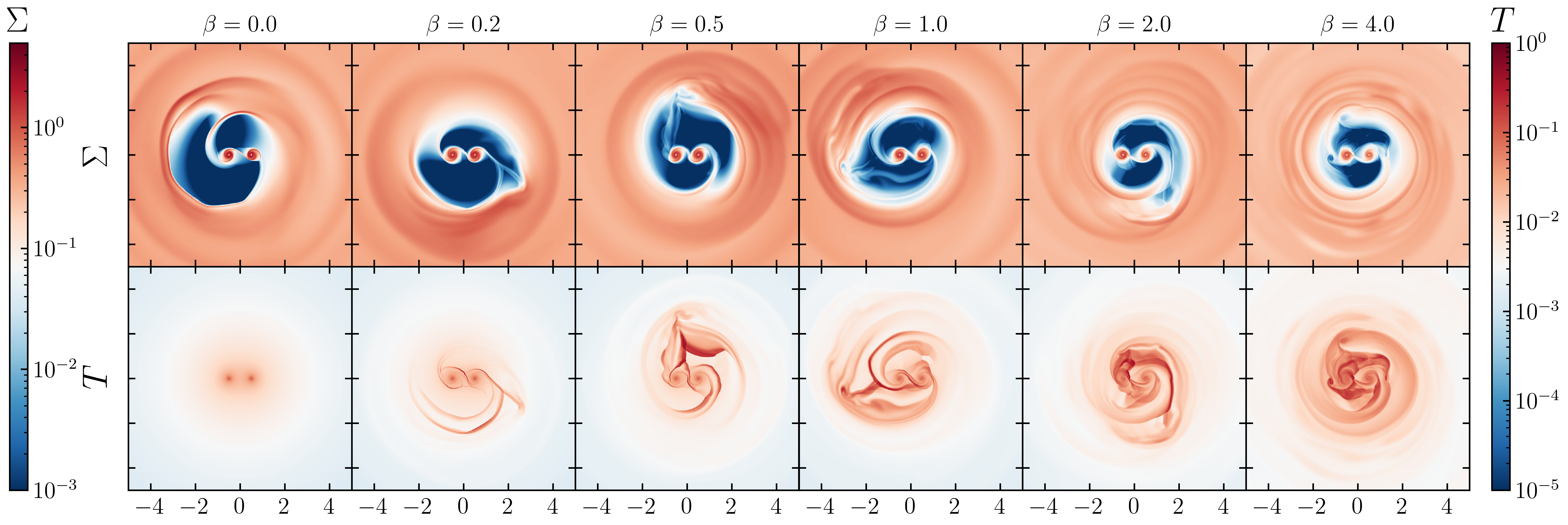}
\end{center}
\vspace{-0.3cm}
\caption{
Similar to Figure \ref{fig:morphology-1}, the density distribution (first row) and the temperature distribution (second row) in parameter survey subset \texttt{t1n3b-}* with fixed $\alpha-$viscosity.
\label{fig:morphology-5}
}
\vspace{-0.3cm}
\end{figure*}

\begin{figure*}
\begin{center}
\includegraphics[width=0.98\textwidth,trim={0cm 0cm 0cm .0cm},clip]{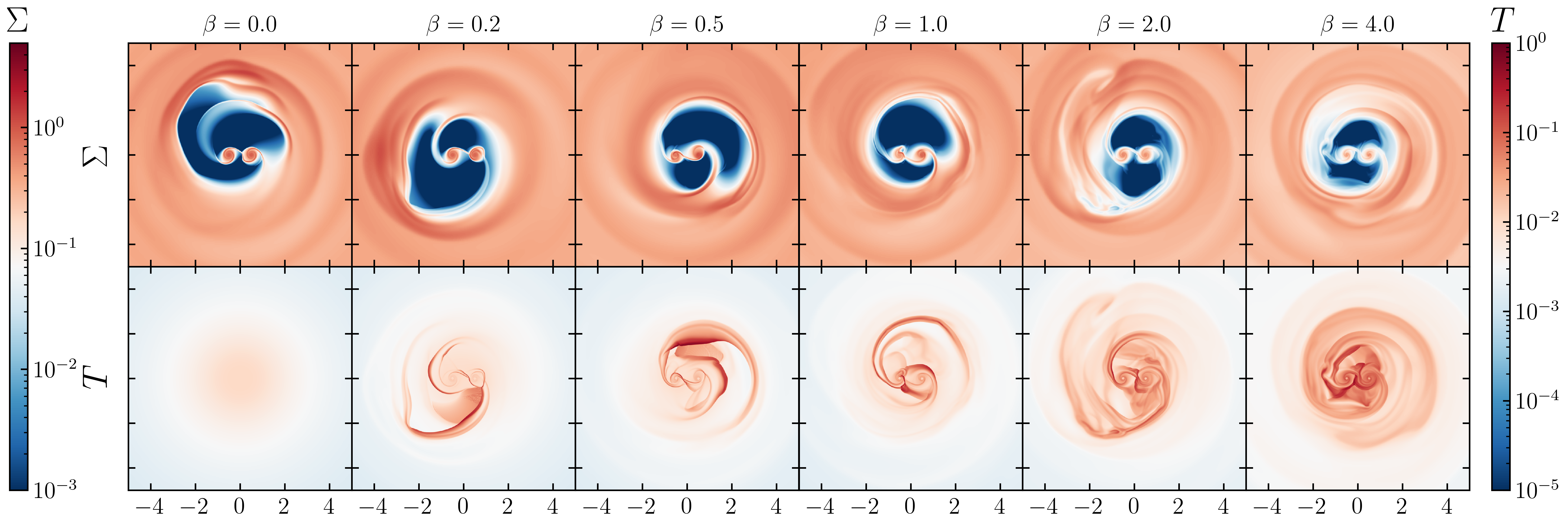}
\end{center}
\vspace{-0.3cm}
\caption{
Similar to Figure \ref{fig:morphology-1}, the density distribution (first row) and the temperature distribution (second row) in parameter study \texttt{t2n2b-}* with fixed $\alpha-$viscosity.
\label{fig:morphology-7}
}
\vspace{-0.3cm}
\end{figure*}

\begin{figure}
\begin{center}
\includegraphics[width=0.45\textwidth,trim={0cm 0cm 0cm .0cm},clip]{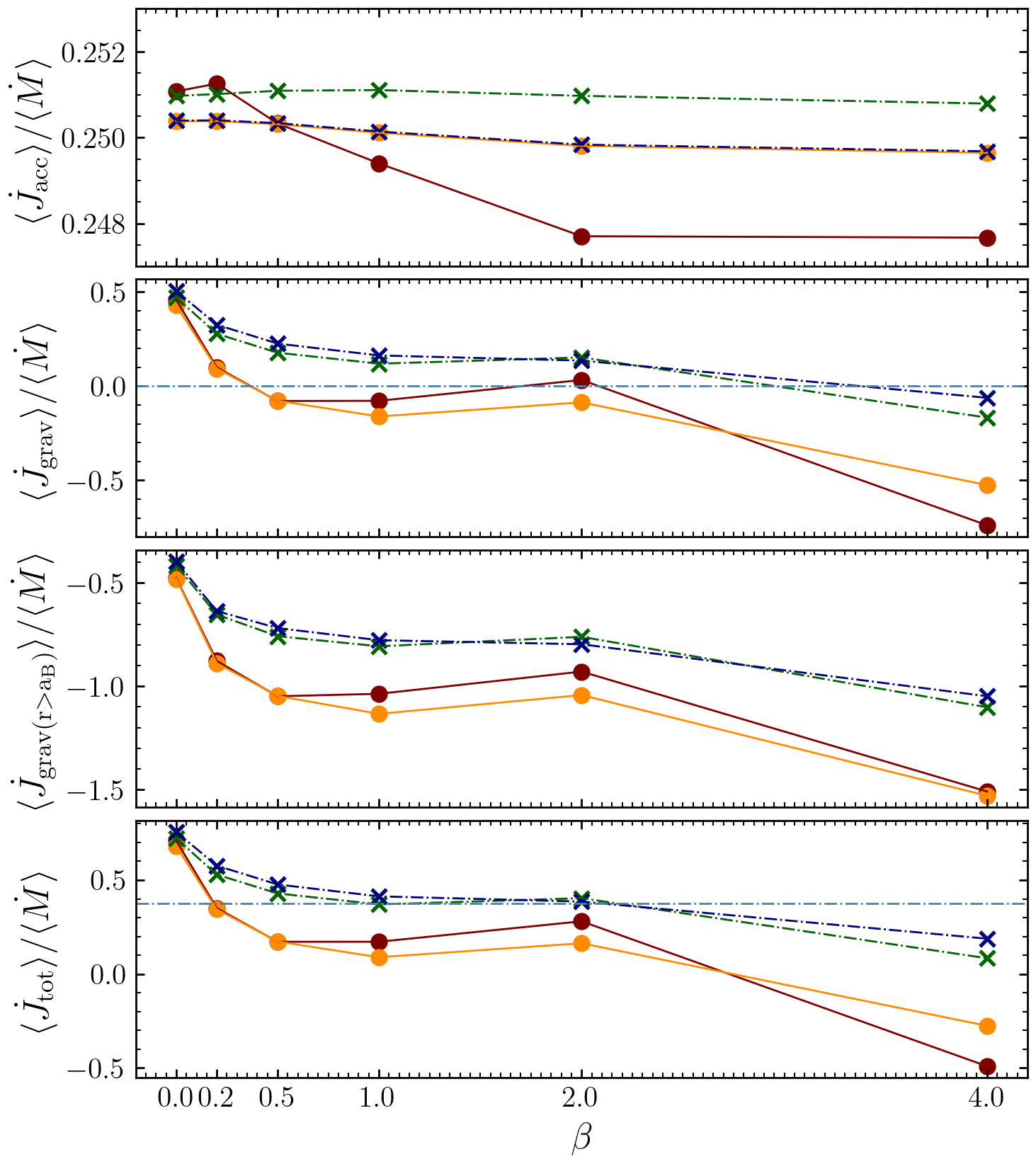}
\end{center}
\vspace{-0.3cm}
\caption{
Similar to Figure \ref{fig:torque-subset1}, various specific torque components and the total specific torque of the CBD acting on the binary as a function of parameterized cooling time $\beta$ in all subset employing a fixed viscosity profile. From the top to the bottom are (1) the specific accretion torque, (2) the specific gravitational torque, (3) the specific gravitational torque from the region $r>a_{\rm{B}}$, and (4) the specific total torque. The horizontal dashed-dotted blue line in the second panel differentiates the sign of total gravitational torque, and the horizontal dashed-dotted line in the fourth panel illustrates the threshold of binary orbital evolution.  
The darkred, orange, darkgreen, and darkblue dots separately represent subset \texttt{t1n1b-}, \texttt{t2n1b-}, \texttt{t1n3b-}*, and \texttt{t2n2b-}*. The solid circles (solid lines) show the runs employing a constant viscosity; while the forks (dash-dotted lines) show the runs employing a fixed $\alpha-$viscosity.
\label{fig:torque-set1}}
\vspace{-0.3cm}
\end{figure}

\begin{figure}
\begin{center}
\includegraphics[width=0.45\textwidth,trim={0cm 0cm 0cm .0cm},clip]{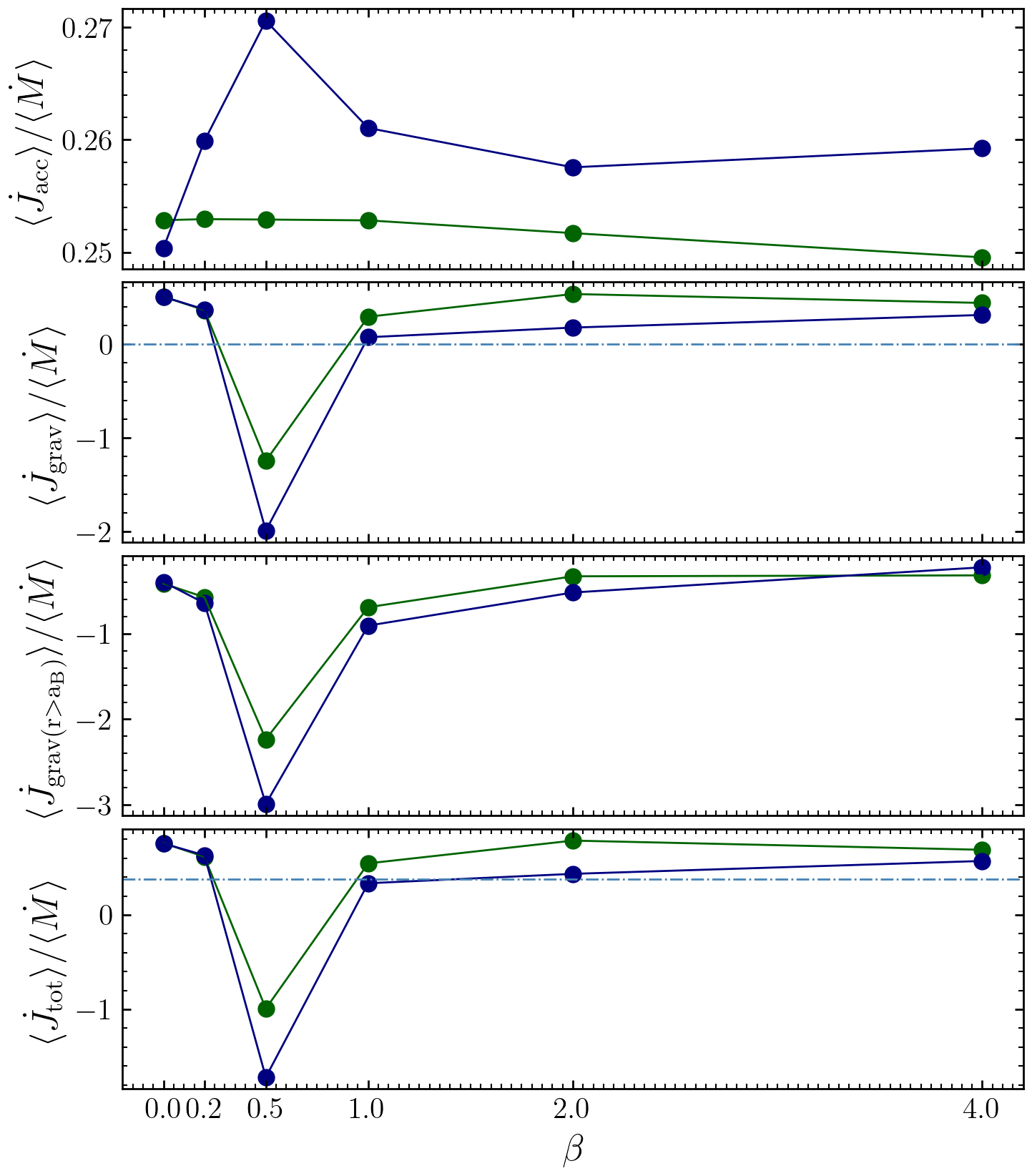}
\end{center}
\vspace{-0.3cm}
\caption{
Similar to Figure \ref{fig:torque-set1}, the darkgreen and darkblue dots separately represent subset \texttt{t1n3b-} and \texttt{t2n2b-}. A dynamically varying $\alpha-$viscosity is adopted in these two subsets. 
\label{fig:torque-set2}}
\vspace{-0.3cm}
\end{figure}

In this appendix, we include the surface density, temperature and viscosity profiles (Figure \ref{fig:morphology-6}, \ref{fig:morphology-4}, \ref{fig:morphology-5}, and \ref{fig:morphology-7}) of all the parameter survey subsets referred in Table \ref{tab:simulation-parameters-maintext}: \texttt{t2n1b-}, \texttt{t2n2b-}, \texttt{t1n3b-}*, and \texttt{t2n2b-}*. Only the simulations in subset \texttt{t2n2b-} are employing viscosity which depends on local temperature. Though having different temperature and viscosity profile comparing to fiducial run subset \texttt{t1n3b-} (detailly discussed in Section \ref{section:results}), a dynamically varying viscosity in subset \texttt{t2t2b-} makes these two subsets of simulations behave similarly: The shape of the cavity first changes from eccentric to circular, then changes to eccentric again with a longer cooling time. The temperature and viscosity in the CBD cavity also increase correspondingly. Simulations from all the other subsets have fixed viscosity profile in the comoving frame of the binary. And the resulting surface density and temperature distributions of these simulations are qualitatively similar: The shape of the CBD cavity generally changes from eccentric to circular as the cooling time increases. The temperature inside the cavity also rises with a longer cooling time. \im{The dependences of various torque components and the total specific torque on cooling parameter $\beta$ are shown in Figure \ref{fig:torque-set1} and Figure \ref{fig:torque-set2}.} 

\bsp	
\label{lastpage}
\end{document}